\documentclass[sigconf]{acmart}
\AtBeginDocument{%
  }

\copyrightyear{2026}
\acmYear{2026}
\setcopyright{cc}
\setcctype{by}
\acmConference[CHI '26]{Proceedings of the 2026 CHI Conference on Human Factors in Computing Systems}{April 13--17, 2026}{Barcelona, Spain}
\acmBooktitle{Proceedings of the 2026 CHI Conference on Human Factors in Computing Systems (CHI '26), April 13--17, 2026, Barcelona, Spain}
\acmPrice{}
\acmDOI{10.1145/3772318.3791022}
\acmISBN{979-8-4007-2278-3/2026/04}

\usepackage[shortlabels]{enumitem}
\usepackage{hyperref}
\usepackage[noabbrev,capitalise,nameinlink]{cleveref}
\usepackage{tabularx}
\usepackage{xcolor, colortbl,booktabs}  
\usepackage[first=0,last=9]{lcg}

\begin{document}

\title[Interaction Data to Empower End-User Decision-Making]{Exploring the Role of Interaction Data to Empower End-User Decision-Making In UI Personalization}


\author{Sérgio Alves}
\email{sfalves@ciencias.ulisboa.pt}
\orcid{0000-0002-3647-0624}
\affiliation{%
  \institution{LASIGE, Faculdade de Ciências, Universidade de Lisboa}
  \city{Lisbon}
  \country{Portugal}
}

\author{Carlos Duarte}
\email{caduarte@fc.ul.pt}
\orcid{0000-0003-1370-5379}
\affiliation{%
  \institution{LASIGE, Faculdade de Ciências, Universidade de Lisboa}
  \city{Lisbon}
  \country{Portugal}
}

\author{Kyle Montague}
\email{kyle.montague@northumbria.ac.uk}
\orcid{0000-0002-9589-9471}
\affiliation{%
    \institution{Computer and Information Sciences, Northumbria University}
    \city{Newcastle upon Tyne}
    \country{United Kingdom}
}

\author{Tiago Guerreiro}
\email{tjguerreiro@ciencias.ulisboa.pt}
\orcid{0000-0002-0333-5792}
\affiliation{%
  \institution{LASIGE, Faculdade de Ciências, Universidade de Lisboa}
  \city{Lisbon}
  \country{Portugal}
}

\renewcommand{\shortauthors}{Alves, S. et al.}

\begin{abstract}
    User interface personalization enhances digital efficiency, usability, and accessibility. However, in user-driven setups, limited support for identifying and evaluating worthwhile opportunities often leads to underuse. We explore a reflexive personalization approach where individuals engage with their digital interaction data to identify meaningful personalization opportunities and benefits. We interviewed 12 participants, using experimental vignettes as design probes to support reflection on different forms of using interaction data to empower decision-making in personalization and the preferred level of system support. We found that people can independently identify personalization opportunities but prefer system support through visual personalization suggestions. Interaction data can shape how users perceive and approach personalization by reinforcing the perceived value of change and data collection, helping them weigh benefits against effort, and increasing the transparency of system suggestions. We discuss opportunities for designing personalization software that raises end-users' agency over interfaces through reflective engagement with their interaction data.
\end{abstract}




\maketitle

\section{Introduction}
User interface (UI) personalization involves tailoring visual and interactive layout elements, such as repositioning or resizing of interface components, to accommodate individual needs, abilities, and usage contexts \citep{paterno2013end, wobbrock2011ability}. The population of users who can benefit from personalization is highly diverse, as it has the potential to enhance accessibility \citep{wu2022reflow,10.1145/3663548.3675650, 10.1145/3613905.3651039}, user experience \citep{nebeling2013crowdadapt}, efficiency \citep{reinecke2011improving}, overall satisfaction \citep{10.1145/3457151, gajos2010automatically, Liu2024}, and website stickiness \citep{10.1080/07421222.2015.1029394}. Personalization can be user-driven \citep{10.1145/3544549.3585668, 10.1145/1166253.1166301}, where individuals manually adjust UI elements, or system-driven \citep{wu2022reflow, bunt2007supporting, KHAMAJ2024164}, where interfaces adapt automatically. Compared to automated approaches, user-driven personalization offers greater empowerment, enhancing users' sense of control and identity \citep{Sundar2010, sundar2008self, marathe2011drives}. However, people often struggle to benefit from it due to uncertainty about their needs \citep{alves}, the effort required to personalize \citep{nngroup, mackay1991triggers}, or the lack of perceived long-term benefits \citep{mackay1991triggers, banovic2012triggering}. This highlights the need to consider personalization solutions that better support users in identifying relevant opportunities, anticipating potential benefits, and implementing changes.

While personalization in general is increasingly shifting toward more system-driven approaches (for example, algorithmic content feeds and recommendations \citep{wu2022reflow, bucher2019algorithmic, montag2019addictive}), where software makes decisions on users' behalf and personalization effort is on the system side, it remains important to consider balanced approaches to UI layout personalization in which users retain control \citep{act2024regulation, valdez2024european}. People, understood here as general users of digital interfaces, often hold contextual knowledge about their routines and needs that designers and automated systems cannot fully anticipate, resulting in personalization needs that only emerge in practice \citep{10.1145/1166253.1166301, 10.1145/1822018.1822019, wobbrock2019situationally}. Consequently, a substantial number of people already modify their interfaces proactively (e.g., through browser extensions \citep{stylish,oldTwitterLayout2025}), representing an active community that has found the need to move beyond default interfaces and personalization.

Supporting these users does not require excluding designers or system involvement. Personalization can remain user-driven even when aided by the system. For example, when the system suggests layout changes, but users retain decision-making control and the ability to further personalize the interface manually. In this context, user-driven personalization aims to empower users and grant them agency over UIs \citep{alves}, allowing them to take action and exert control over the interfaces they use. However, can these users be fully empowered if they lack the knowledge and data to make informed personalization decisions based on actual usage rather than perceptions? This question highlights a gap we address by studying balanced alternatives to system-driven solutions.

In this paper, we explore the role of interaction data\footnote{``Interaction data'' refers to recorded information (i.e., logs) about how users engage with an interface, including actions such as clicks, mouse movements, scrolls, and keystrokes \citep{10.1007/978-3-031-16103-2_7}.} in supporting users' data-driven decision-making in UI personalization. We focus on efficiency-related data (e.g., click activity), as efficiency is users' primary personalization goal \citep{reinecke2011improving, alves}. Although companies frequently collect such data to optimize their products, using tools like \textit{Google Analytics} \citep{Analytics}, it typically has been inaccessible to end-users. In personalization contexts, interaction data is often translated into high-level recommendations \citep{wu2022reflow, gajos2017influence} rather than presented directly to users, although many expect recommendations with metrics or graphical data representations \citep{bunt2007understanding}. To the best of our knowledge, the potential of making interaction data available to personalization users has not yet been explored. Such data can empower users to identify opportunities for user-driven personalization and to critically assess system-initiated adaptations.

With the ultimate goal of informing the design of personalization software that empowers users to make informed, data-driven, decisions, we defined the following research questions (RQs):

\begin{enumerate}
    \item \textit{How do people engage with interaction data to identify and reflect on opportunities for UI personalization?}
    \item \textit{What are people's perceptions of the processes of collecting and analyzing interaction data for data-driven personalization, including its challenges and benefits?}
    \item \textit{How do people envision personalization solutions enabling access to interaction data, and what support do they expect for identifying opportunities, making decisions, and implementing changes?}
\end{enumerate}

\begin{figure*}[h]
    \centering
    \includegraphics[width=\textwidth]{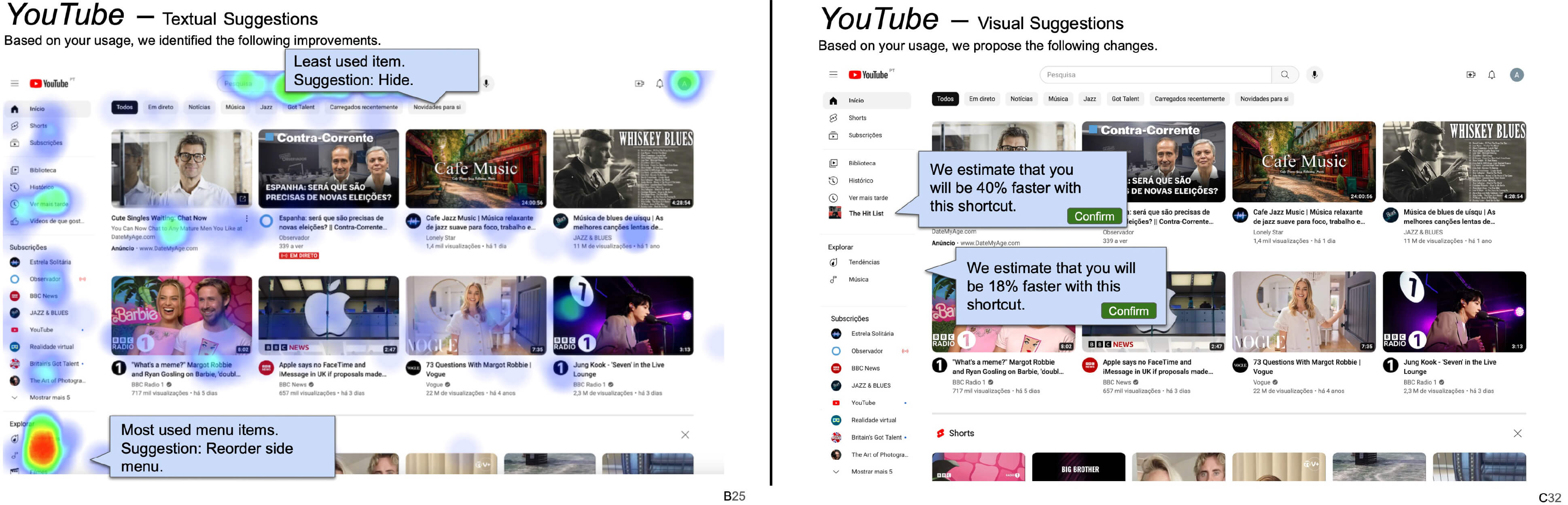}
    \caption{Two example vignettes illustrating two of the four scenarios in which a hypothetical platform leverages personal interaction data to support the identification and implementation of UI personalization opportunities. Each scenario was illustrated through multiple vignettes, which served as design probes to provoke participant reflection and discussion during the interviews. In these two illustrations, the hypothetical software supports decision-making by generating textual (left) or visual (right) personalization suggestions based on clickstream data. The left vignette shows the original UI overlaid with a click heatmap and textual suggestions (interpretations of the interaction data that highlight potential usability issues and recommend manual personalization actions). The right vignette, through a system-initiated design process, presents visual suggestions, where the system displays a preview of the personalized UI, which users can adopt and further refine, and includes estimated time savings. Similar benefit estimates are also present in other vignette scenarios. Both scenarios illustrate a workflow where users retain personalization control and can personalize their interfaces freely, following a user-driven strategy.}
    \label{fig:previewYT}
    \Description{Two side-by-side screenshots of the YouTube homepage showing textual and visual personalization suggestions. The left screenshot includes a heatmap overlaying YouTube and two pop-ups; the right screenshot shows a modified YouTube homepage with a pop-up indicating estimated time savings.}
\end{figure*}

We conducted a semi-structured interview study with 12 participants, employing experimental vignettes as design probes (\cref{fig:previewYT}) to investigate people's responses to different ways of engaging with interaction data to support UI personalization. The vignettes depicted interfaces from a hypothetical platform that enables interaction data collection and visualization to support smartphone and computer personalization. We divided the vignettes into four sets, each illustrating different methods of presenting and using the data to guide personalization: \textit{A: Data-Only Visualization Dashboards} (no support for interpreting data or personalizing), \textit{B: Textual Suggestions} (\cref{fig:previewYT}, left), \textit{C: Visual Suggestions} (\cref{fig:previewYT}, right), and \textit{D: Social Suggestions} (recommendations informed by users with similar behaviors \citep{Hong27082025}). The vignettes, developed solely for the interviews, used synthetic data and familiar UIs to help participants engage with the data and reflect on different levels of support for decision-making and personalization. The discussions also explored participants' reactions to system-initiated design features (\textit{Visual Suggestions}), cost-benefit trade-offs highlighting effort and time savings, and data-sharing comfort (\textit{Social Suggestions}).

Results indicate that most participants can identify personalization opportunities by relying solely on interaction data, with click heatmaps being especially effective (\cref{fig:previewYT}, left). Nevertheless, most favor a personalization workflow with a system-initiated design process supported by visual suggestions  (\cref{fig:previewYT}, right), valuing interaction data visualizations to understand the suggestions and justify their decisions. Participants noted that access to interaction data can enhance transparency and trust in personalization software, supporting personal informed choices in dubious situations and potentially increasing their willingness to share data. We also found that accessing the data behind personalization suggestions can be important for enhancing the perceived value of personalization, motivating users to reflect on their interaction patterns and take action. Ultimately, these findings suggest that users feel most empowered in personalization when they not only have access to interaction data but are also offered actionable personalization options they can further adjust, supported by mechanisms that help them determine when and how to personalize.

We contribute design considerations to inform future solutions that leverage interaction data to support user-driven personalization. Specifically, we discuss strategies for integrating interaction data into software that (1) help users identify personalization opportunities, (2) streamline the design and implementation process, (3) highlight both the perceived and practical benefits (such as increased efficiency and usability), and (4) empower individuals to confidently take control of their UIs, ultimately fostering adoption and long-term engagement.

\section{Related Work}
UI personalization is a response to the challenge of creating products accessible to the majority, an ambition shared by paradigms like \textit{inclusive design} \citep{clarkson2004countering, keates2000towards} or \textit{ability-based design} \citep{wobbrock2011ability,10.1145/3676524,mitchell2024physiological,10.1145/3706598.3713367}. However, these paradigms are often considered utopian due to the difficulty of designing a universally usable UI \cite{newell2000user}. Even automatically generated UIs (e.g., SUPPLE \cite{gajos2010automatically}), tailored to individuals' abilities or context, might not always align perfectly with the multitude of users' preferences or the dynamics of human behavior. This makes UI personalization crucial for enabling real-time adjustments in response to user characteristics or usage contexts unanticipated by the original designers, such as situationally-induced impairments \citep{wobbrock2019situationally}. We reviewed existing work on UI personalization and digital approaches to empower individuals with their personal data.

\subsection{Personalization of User Interfaces}
Personalization mechanisms can be either built into interactive systems (e.g., color modes) or introduced as third-party repair solutions (e.g., browser extensions) \citep{paterno2013end}. Built-in user-driven approaches include adaptable interfaces and menus that allow users to perform predefined configurations \cite{10.1145/1166253.1166301, 10.1145/985692.985704}. On the system-driven end, built-in solutions include adaptive \cite{kuhme1993user, jameson2007adaptive, Liu2024} and context-aware UIs \cite{10.1007/978-3-540-39653-6_12}, which alter their appearance in response to users' behavior or contextual factors. These solutions are often part of broader concepts such as intelligent user interfaces \citep{10.1007/978-981-97-5810-4_24, 10.1145/3377325.3377500, 10.1145/238218.238323} or intent-based user interfaces \citep{ding2024intentbaseduserinterfacescharting}. Despite these developments, most user interfaces remain static or lack the flexibility to accommodate known personalization needs, such as the ability to reposition interface elements~\citep{nebeling2013crowdadapt, alves}. In practice, built-in UI personalization is generally limited to changing color schemes, font sizes, or brightness. 

Third-party runtime repair solutions aim to address these limitations through system-driven adaptations, offering cultural adaptivity \citep{reinecke2011improving}, example-based adaptation \citep{kumar2011bricolage}, or UI optimization based on task performance prediction \citep{duan2020optimizing}; and user-driven customizations, including solutions like \textit{GitUI} \citep{alves, 10.1145/3544549.3585668}, \textit{Pagetailor} \citep{bila2007pagetailor}, and \textit{Stylette} \citep{10.1145/3491102.3501931}. These solutions provide users with personalization options when the original UI lacks intelligence or fails to incorporate personalization. For example, \textit{Reflow} is an adaptation software \cite{wu2022reflow} that extracts the layout of mobile app screens, refines it using machine learning, and re-renders the personalized UI based on usage data. In contrast, \textit{CrowdAdapt} \citep{nebeling2013crowdadapt} allows the manual customization of websites through operations like move, resize, and hide.

Recent advances in large language models (LLMs) and conversational agents are also advancing the forms of personalization. UI layouts and code can now be generated or modified through natural language \citep{10903712}, leading to concepts such as vibe coding \citep{meske2025vibe, 10.1145/3715336.3735766}. This has expanded the population capable of creating or modifying software, resulting in the rise of ``citizen developers'', non-professional developers using low-code and no-code tools \citep{lugovsky2021lowcode, muhammad2024citizen}.

In the context of UI personalization, most of these solutions (e.g., \textit{Dynavis} \citep{vaithilingam2024dynavis} or \textit{NLDesign} \citep{zhang2024nldesign}) focus on design-time support. For example, generative and malleable UIs \citep{cao2025generative} aim to produce layouts dynamically in response to user prompts and continuously adapt them as needs evolve. Conversely, a smaller number of solutions focus on runtime UI repair, enabling personalization in systems whose original designs are not inherently malleable \citep{li2023using, 10.1145/3491102.3501931}. Despite progresses in modeling UIs \citep{haddad2024good, bai2021uibert}, design preferences \citep{BrandenburgerJanneck2024}, or individual perception of UIs \citep{micheli2024modeling, peng2025designpref}, these approaches can struggle with vague instructions \citep{10.1145/3491102.3501931}, accounting for unique individual preferences \citep{BrandenburgerJanneck2024}, and translating natural language into actionable commands \citep{li2023using}. Furthermore, like other user-driven personalization solutions, they are constrained by users' difficulty in understanding their needs and the value of personalizing \citep{alves,mackay1991triggers, banovic2012triggering}.

The advantages of user- and system-driven personalization relate to the control people desire and the effort and privacy they are willing to provide, creating cost-benefit trade-offs associated with these factors \citep{10.1007/978-3-642-40483-2_44, Sundar2010}. User-driven approaches do not require personal data collection and provide a higher sense of control and identity \cite{sundar2008self, marathe2011drives}. However, users need to invest significant time and effort, which may outweigh personalization benefits \citep{bunt2007supporting} and result in a tedious personalization process \citep{10.1145/3491102.3501931}. Conversely, while system-driven personalization minimizes user effort, it diminishes the sense of control, raises privacy concerns due to the extensive data collection required by these processes \citep{Sundar2010}, and increases cognitive load by disrupting established mental models when the UI changes \citep{10.1007/978-3-642-40483-2_44, 10.5555/3033040}.

Overall, users who prefer system-driven approaches tend to value the outcomes (i.e.,  the personalized UI) while those who favor user-driven approaches are more engaged with the process itself, often driven by a desire for agency \citep{Sundar2010}.

\subsection{Managing User-driven Personalization Trade-offs}
This work explores ways of increasing users' agency over both their interfaces and the personalization process, making user-driven personalization (and its associated trade-offs) a central concern. Several alternatives have emerged to balance the effort of user-driven personalization with the control it offers over system-driven solutions. These include example-based personalization \citep{kumar2011bricolage, 10.1145/1753326.1753667, fitzgerald2008copystyler}, which lowers effort by allowing users to transfer design between UIs, programming-by-example \citep{10.1145/3640543.3645176}, and social or collaborative approaches \citep{10.1145/3544549.3585668, alves}, which distribute the implementation effort across a community. The most prevalent strategy involves mixed-initiative personalization \citep{debevc1996design}, in which systems like \textit{UIFlex} \citep{10.1145/3457151} or \textit{MICA} \citep{bunt2007supporting} offer suggestions that users can accept or reject.

Even with reduced effort, the success of user-driven personalization still hinges on users perceiving its value \citep{alves}. To support a fair cost-benefit analysis, it is crucial to make users aware of potential benefits, as they are unlikely to invest time personalizing unless they find it worthwhile \citep{mackay1991triggers}. This includes helping them recognize opportunities, estimate the potential return on their effort (i.e., benefits), and visually demonstrate how much better their UIs could be through personalization, such as by previewing changes \citep{alves, 10.1145/3491102.3501931}.

Overall, most effort-reduction solutions shift control over personalization details to the system, often sacrificing user control over the process and the result. In this work, we address cost-benefit trade-offs from both sides, studying balanced forms of alleviating the cost of identifying personalization opportunities while raising awareness of its benefits. We explore for the first time, to the best of our knowledge, how accessing personal interaction data can foster and improve user-driven personalization by (1) supporting self-reflection on behavior, (2) revealing unanticipated personalization opportunities, such as interaction inefficiencies, and (3) enabling users to better estimate the cost-benefit trade-offs of UI personalization.

\subsection{Interaction Data to Enhance the Digital Experience}
Personal interaction data has been used to improve various types of personalization and digital experiences. In UI personalization, system-driven methods frequently leverage this data to create spatial maps \citep{wu2022reflow} or apply recency- and frequency-based algorithms for UI adjustments \citep{gajos2017influence}, which raises the need to increase the transparency of data usage and system decisions \citep{10.1145/1357054.1357252}. Conversely, user-driven approaches rarely collect personal interaction data, while mixed-initiative alternatives typically rely on user models predicting performance or actions from user characteristics \citep{10.1145/3457151, bunt2007supporting}. Mixed-initiative solutions often provide suggestion rationales focused on system operation and estimated time savings, but rarely explain how users' data informs suggestions. Users have found the savings unclear or minimal and expected rationales to include personalized information and graphical data \citep{bunt2007understanding, bunt2009mixed}.

Interaction data powers content personalization (e.g., targeted advertisement \citep{lina2021privacy}, recommender systems \citep{7918058}, personalized search \citep{toch2012personalization}, and marketing \citep{PLAZA2011477}) and informs iterative UI design, helping designers refine interfaces based on real user behavior \citep{muresan2009integrated}. Development teams typically use web analytics tools like \textit{Google Analytics} \citep{ledford2011google} to track user tasks and challenges by aggregating metrics like visitor counts, page views, and session duration. Complementary methods involve capturing UI events (e.g., clicks \citep{kaur2015analysis}), failures or unexpected situations \citep{harty2021logging}, and server-side traffic logs~\citep{hong2001webquilt}.

In summary, while these uses of interaction data can indirectly benefit end-users, they rarely translate into a feeling of empowerment or control \citep{10.1007/978-3-540-39653-6_12, auxier2019americans}, with the data subjects (the end-users themselves) perceiving such data practices as posing more risks than benefits \citep{auxier2019americans}. Individuals' access to personal interaction data is mainly limited to applications or operating systems offering access to screen time data \citep{screentime} or usage history (e.g., browser history). This gap between data collection and user control has led to concepts such as \textit{data-owning democracy} \citep{doi:10.1177/14748851221110316} or research fields like \textit{Human–Data Interaction} \citep{mortier2014human, niu2025chat, rey2024understanding}, which emphasize the importance of giving individuals more meaningful ways to engage with their data. Building on this context, and considering users' challenges in identifying when and how to personalize \citep{alves} and in noticing the benefits of change \citep{mackay1991triggers, banovic2012triggering}, we study a novel approach that offers users an accessible way to engage with their interaction data. This approach aims to foster reflection and empower users to make independent, transparent, and informed decisions about the UIs they interact with and themselves.

\section{Materials and Methods}
We conducted a qualitative semi-interview study employing experimental vignettes as design probes to elicit participants' perspectives on how they might engage with interaction data to identify and reflect on personalization opportunities. 

The study used 42 vignettes, divided into four sets (\textit{A} -- \textit{D}), each illustrating a different version of the hypothetical platform, \textit{UIPulse}. These versions were intentionally designed to represent different scenarios for supporting interaction data visualization and personalization, thereby provoking discussions around dimensions such as effort, privacy, and control. Each of the four sets follows a structured format: beginning with high-fidelity illustrations of the corresponding scenario and concluding with a reflection moment, a series of elicitation questions used to guide the semi-structured interview and capture participants' perspectives.

Our university's ethics committee reviewed and approved the research protocol. This section details our methodology, including the vignette development and study procedure.

\subsection{Goals and Methodological Approach}
This study explored how users perceive the value of engaging with interaction data, not simply to automate personalization but to reflect on behaviors and identify UI improvements. We focused on the ideation and reflection processes rather than evaluating real-world task performance or efficiency gains, which are well documented in previous research \citep{wu2022reflow, reinecke2011improving}. We investigated how participants envisioned the collection and use of interaction data, and whether they could move from passive viewing to identifying actionable personalization opportunities. To support this open-ended reflection and inform future design directions, we used speculative scenarios as design probes instead of a functional system.

\subsubsection{Vignettes as Design Probes}
Design probes are speculative artifacts or scenarios used in the design process to provoke reflection and discussion about future technologies or practices \citep{10.1145/2470654.2466473}. They help explore participant perceptions in unfamiliar or yet-to-exist contexts before developers commit to a specific design direction.

We used on-paper vignettes as design probes. Vignettes are short scenarios involving personas, objects, interfaces, or contexts of use (e.g., a story) and are well-established in qualitative research for eliciting in-depth perspectives and judgments \citep{jmmss833, finch1987vignette, baguley2022statistical}. Vignettes are commonly used alongside other data collection methods, such as interviews, observations, or surveys. We combined vignettes with interviews, an effective method for prompting discussion and contextualizing participant perspectives \citep{kandemir2018using, barter2000wanna, barter1999use}.

Compared to using functional systems, vignettes are more cost-effective, time-efficient, and flexible. They enable the discussion of multiple imaginative futuristic alternative solutions to the same problem without the constraints or bias of technical implementation. To support this, vignettes must provide sufficient context about the topic under discussion while remaining vague enough for participants to fill in the missing details \citep{Hypotheticallyspeaking, doi:10.1177/16094069221074495, barter1999use, barter2000wanna}.

Vignettes can also ease recruitment by reducing confidentiality and privacy concerns present in more intrusive methods \citep{doi:10.1177/16094069221074495}. Rather than collecting participants' authentic interaction data, we constructed vignettes with synthetic data originating from UIs that participants were familiar with. This trade-off grounded discussion in familiar contexts while preserving privacy. While authentic interaction data might offer richer detail, synthetic examples can still enable valuable insights into how participants interpret data and consider personalization opportunities.

When the topic is context-dependent, vignettes present concrete examples of people and their behavior \citep{hazel1995elicitation}. Participants are then asked to respond based on what they or someone else might do in a given situation. To support this approach, we introduced two personas (Alex and Chris) to provide narrative and contextual grounding for how data could be collected and used \citep{hughes1998considering}. This allowed us to frame synthetic data as belonging to the personas, making it easier to illustrate and discuss interaction data in scenarios where participants might have been reluctant to share real information \citep{Sundar2010}, such as those involving social media platforms.

We were interested in studying how people interpret the data and make personalization decisions, which vignettes are well-suited to explore \citep{finch1987vignette}. While participants' responses may not necessarily reflect real-world behavior \citep{barter1999use}, this is less of a concern given our focus on studying interpretation and decision-making \citep{finch1987vignette}.

\subsubsection{Scenarios Selection}\label{ScenariosSelection}
The four illustrated \textit{UIPulse} versions were developed to differ in their use of interaction data to support personalization, aiming to provoke reflection on existing trade-offs and the practical limits of system-side support. The scenarios vary along key design dimensions known to influence engagement with personalization systems, namely user effort \citep{bunt2007supporting, 10.1007/978-3-642-40483-2_44}, privacy \citep{Sundar2010}, sense of identity \citep{marathe2011drives}, and control \citep{sundar2008self, 10.1007/978-3-642-40483-2_44, marathe2011drives}. These dimensions can be particularly connected to how interaction data is utilized for personalization: while greater system support in data translation and analysis can reduce effort, it may simultaneously decrease perceived control or raise privacy concerns.

Prior personalization work balances these dimensions by offering suggestions \citep{bunt2007understanding, bunt2009mixed} or social support through community-based templates \citep{alves, 10.1145/3544549.3585668}. In our vignettes, we reframed these strategies from a user-driven perspective, aligning with a more balanced approach. Instead of focusing on how system-driven approaches could benefit from interaction data (e.g., improved explainability of system decisions), we reimagined them as ways to support user decision-making within a user-driven personalization process informed by interaction data. For example, in a low-effort condition (visual suggestions), the vignettes present a \textit{UIPulse} version that guides users through the interpretation of data and the initiation of personalization actions. This represents a user-driven personalization strategy guided by system-initiated design considerations, which lowers the effort and control and requires greater data analysis (potentially increasing privacy and lack of control concerns). \cref{table:UIPulseSets} summarizes key differences across scenarios.

\begin{table*}[h]
\centering
\begin{tabular}{p{7cm}|c|c|c|c} % first column fixed width
\toprule
\textbf{Scenario} & \textbf{\begin{tabular}{@{} c @{}}Personalization \\Control\end{tabular}} & \textbf{\begin{tabular}{@{} c @{}}Design \\Initiative\end{tabular}} & \textbf{\begin{tabular}{@{} c @{}}Cost–Benefit\\ Estimates\end{tabular}} & \textbf{\begin{tabular}{@{} c @{}}Data Analysis\\ Scope\end{tabular}} \\
\midrule
\textbf{Base Visualization Dashboards (\textit{A})} \newline \small Introduces screen time and clickstream data visualizations for two applications.
& User-driven & User-initiated & — & None (Data only)\\
\midrule
\textbf{Base \& Textual Suggestions (\textit{A} + \textit{B})} \newline \small Adds textual personalization suggestions with rationale and cost-benefit estimates.
& User-driven & User-initiated & \checkmark & Local\\
\midrule
\textbf{Base \& Visual Suggestions (\textit{A} + \textit{C})} \newline \small Adds visual UI modifications that users can accept.
& User-driven & System-initiated & \checkmark & Local\\
\midrule
\textbf{Base \& Social Suggestions (\textit{A} + \textit{D})} \newline \small Adds socially-informed UI modifications that users can accept.
& User-driven & System-initiated & \checkmark & Socially-Informed\\
\bottomrule
\end{tabular}
\caption{\textit{UIPulse} scenarios. Across all scenarios, users control the personalization process, with access to the base data-only dashboards (set \textit{A}) and a manual UI customization software. The scenarios differ regarding design initiative, the presence of cost-benefit estimates (e.g., time-saving trade-offs), and the scope of data analysis. \textit{B} supports only the interpretation of data through textual suggestions; \textit{C} and \textit{D} support both interpretation and execution via pre-configured UI changes (i.e., system-initiated design). \textit{A}--\textit{C} provide tailored solutions unique to each user, while \textit{D} introduces socially-informed suggestions based on behavioral patterns shared across users.}
\Description{A table summarizing the four UIPulse scenarios (A–D) across personalization control, design initiative, cost–benefit estimates ("—" means no estimates are present), and data-analysis scope.}
\label{table:UIPulseSets}
\end{table*}

\subsection{Procedure}
We conducted individual semi-structured interviews supported by the printed vignettes. Each session took place in a private room at the university and lasted on average 93.8±27.2 minutes, with short breaks at key moments.

We first collected informed consent and explained that the purpose of the session was to understand their perception of an interaction data collection and visualization platform. Participants were not intentionally debriefed on the study's focus on UI personalization. Instead, we broadly described the goal as exploring potential uses of interaction data. This allowed us to examine whether participants would intuitively recognize UI personalization as a meaningful use case for the data or propose alternative applications.

We then handed participants the first set of vignettes. We asked them to read the vignettes and express their thoughts. We handed out the sets of vignettes one at a time, with participants unaware of the next set's content, as there was a feature dependency between sets. We wanted participants to experience the personalization features progressively, allowing them to share any unconventional opportunities for using interaction data before being exposed to the additional capabilities introduced in later sets. After the first set of questions about set \textit{A}, we introduced the notion of personalization, which guided subsequent discussions. Participants consulted the vignettes fluidly, often spending less than a minute on each and revisiting them as needed throughout the session.

The reflection moments (elicitation questions) embedded in the vignette structure served as prompts to guide the semi-structured interviews. During each moment, the researcher read aloud and explained the associated questions, encouraged open discussion, and asked follow-up questions based on participants' responses. Throughout the session, the researcher also paid attention to participants' comments and body language to probe further into their reasoning and perspectives.

\subsection{Vignette Design}
\begin{figure*}[h]
    \centering
    \includegraphics[width=0.80\linewidth]{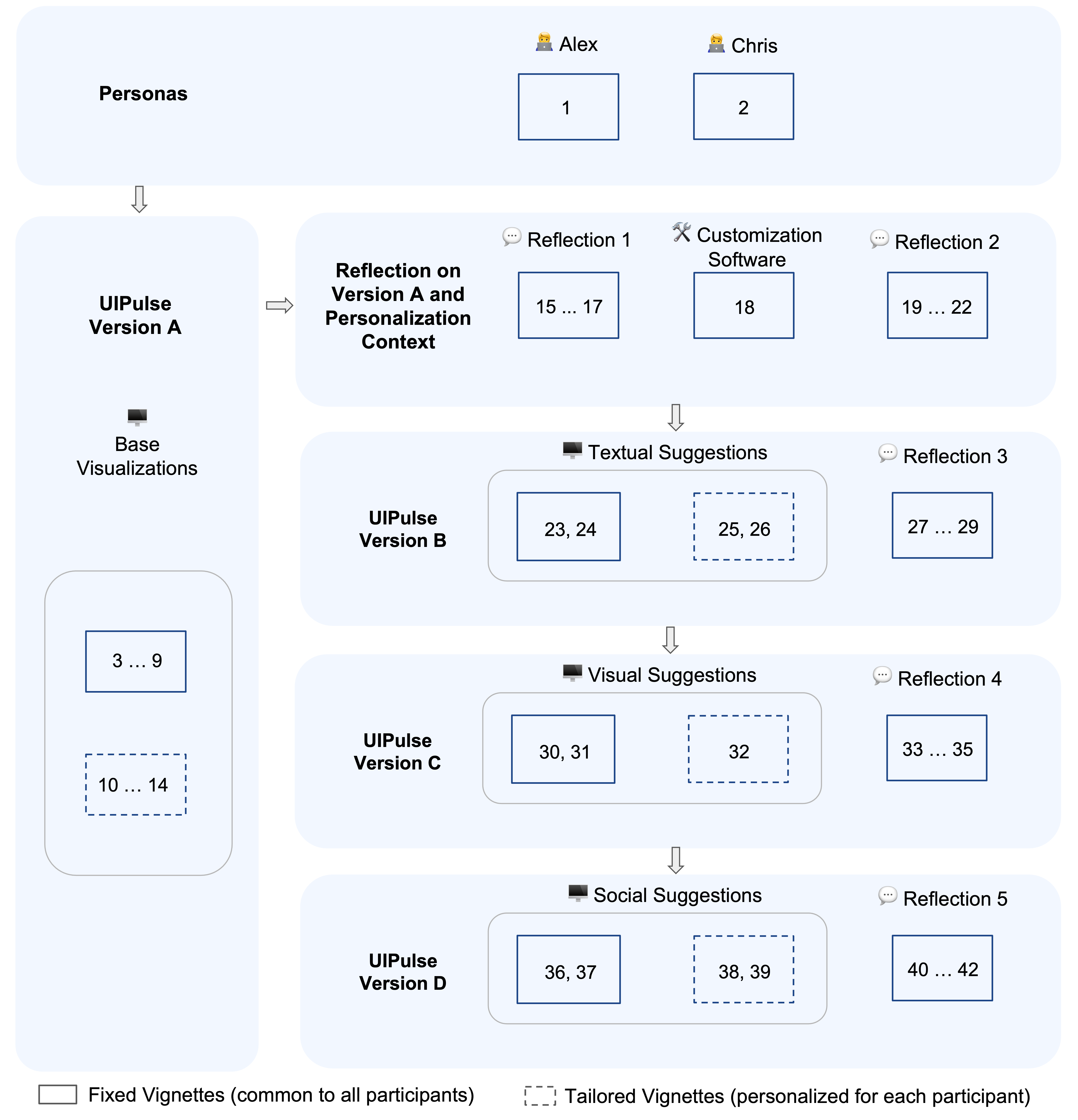}
    \caption{Structure of the four vignette sets used to support the interviews, where each set represents a different version of UIPulse. Each set consists of a series of vignettes, with their ranges indicated using ellipses (e.g., 15...17). The sequence begins with the presentation of two personas, followed by Set \textit{A}, which introduces a UIPulse version focused on data visualization dashboards (accessible across all sets). After reflecting on these dashboards, participants are introduced to the concept of personalization through a vignette illustrating a customization tool and invited to consider how interaction data might support personalization. Sets \textit{B}, \textit{C}, and \textit{D} follow a similar structure, each featuring UIPulse illustrations that include both fixed vignettes and vignettes tailored individually for each participant, along with a reflection moment.}
    \Description{Illustration showing six connected containers representing the personas and UIPulse vignette sets (versions A–D) with reflection moments. The Personas container includes Alex and Chris and connects to Version A, which has fixed and tailored vignettes followed by two reflection moments. Versions B, C, and D are stacked similarly, illustrating textual, visual, and social suggestions (through fixed and tailored vignettes), each followed by a reflection moment.}
    \label{fig:vignettesdiagram}
\end{figure*}

The vignettes are structured to first present the story of Alex and Chris. They are then divided into four sets, with each set presenting a different \textit{UIPulse} version and concluding with the corresponding reflection moment. Each version was represented with interaction data from two applications: one fixed web application shown to all participants, and another individually selected web or mobile application tailored to each participant. \cref{fig:vignettesdiagram} illustrates the structure of our vignettes, which were informed by the work of \citet{Hypotheticallyspeaking}, including the development of personas and associated questions. The complete set of vignettes and the interview script are available online\footnote{https://osf.io/8awdj/?view\_only=9451ed14653b4a96857750ff9eeb9b3b}. We provide a sample of these vignettes in \cref{appendixbignettes}.

A pilot study involving two participants (P1 and P2) helped refine the materials by reducing the number of vignettes and questions to minimize cognitive load and potential participant fatigue.

\subsubsection{Personas}
We introduced two personas to ensure participants could relate to at least one, helping them engage with the context and feel comfortable expressing critical views. The names Alex and Chris were intentionally chosen because they are commonly used as gender-ambiguous names. The personas differ in age, education, profession, digital literacy, and intention to use \textit{UIPulse}. Alex is a 32-year-old tech-savvy user. Regularly engaging with digital devices for work and personal use, Alex has become curious about their interaction data and installs \textit{UIPulse} to explore it. In contrast, Chris, 59 years old, has limited digital experience. Chris recently discovered \textit{UIPulse} and decided to access it for the first time.

\subsubsection{Designing Vignettes with Data from One Common and One Participant-Specific Application} 

To help participants ground their reflections in familiar contexts, each vignette set included vignettes with interaction data from two applications: one fixed across all participants, and one tailored to each individual.

The fixed application was the website of an international retail chain\footnote{https://www.lidl.com}, which allows people to browse products and consult discount flyers. We included data from this application to ensure that at least one source provided sufficiently complex and engaging interaction data.

We selected the tailored application based on a preliminary questionnaire, where participants listed three applications or websites they frequently used. From these, we chose the one most likely to provide meaningful interaction patterns and personalization opportunities, often a social media, news, or entertainment platform like \textit{YouTube} (\cref{fig:previewYT}).

We defined specific vignette slots for presenting data from both applications, ensuring a consistent placement for all participants (as shown in \cref{fig:vignettesdiagram}).

\subsubsection{High-fidelity Illustrations of UIPulse}
We illustrated the four scenarios (\cref{ScenariosSelection}) based on relevant related work and popular commercial tools:

    \begin{figure*}[h]
    	\centering
    	\includegraphics[width=1\linewidth]{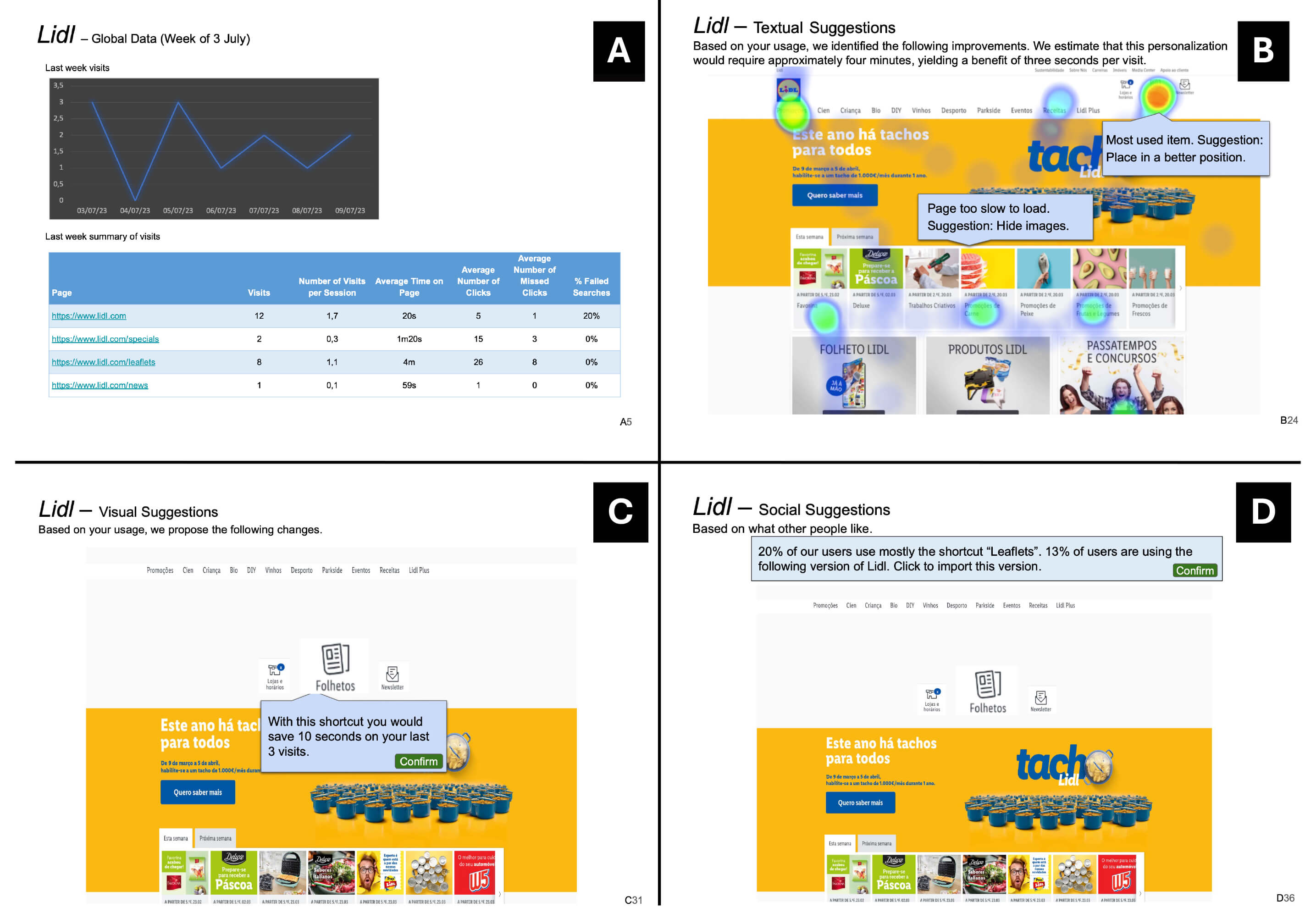}
    	\caption{Example vignettes from each of the four \textit{UIPulse} sets. Each vignette includes a title (combining the vignette name with the title of the analyzed UI) and a brief description.} %To see the details of each category, please check the Supplementary Materials (Sup-D).
    	\label{fig:vignettes4}
        \Description{Four example vignettes from the UIPulse sets showing different personalization approaches for Lidl.com. (A) Base data-only visualizations with charts and tables of interaction metrics. (B) Textual suggestions overlaying a heatmap with pop-ups recommending interface changes. (C) Visual suggestions with pre-configured UI modifications and time-saving estimates. (D) Social suggestions showing community-informed recommendations with pop-ups for importing alternative layouts.}
    \end{figure*}

\begin{enumerate}[(A)]
\item \textit{Base Data-Only Visualization Dashboards.}
    Set \textit{A} begins with screen time data (inspired by \textit{macOS Screen Time} \citep{screentime}) to introduce participants to Alex and Chris. It then presents clickstream data for two selected applications, using both numerical metrics (e.g., visit counts, average time per UI; \cref{fig:vignettes4} A) and visualizations like click and scroll heatmaps (similar to \cref{fig:previewYT}, left), highlighting frequently used UI elements and sections.

    These data representations are available across all \textit{UIPulse} sets, enabling users to consult them regardless of the scenario. The dashboards follow standard practices for visualizing usability \citep{usabilitymetrics}, clickstream \citep{7918058, CrazyEgg}, and traffic data \citep{Analytics, Piwik}. Heatmaps were generated using \textit{ClickHeat} \citep{Clickheat}, a software tool that models the functionality of commercial platforms such as \textit{CrazyEgg} \citep{CrazyEgg}.

    Additionally, to clarify the concept of personalization, this set includes an illustration of a customization tool that enables users to reposition interface elements or adjust their visual properties (e.g., color, size).% This aims to clarify the concept of personalization by providing concrete examples of actionable interface modifications.

    \item \textit{Textual Personalization Suggestions.} Set \textit{A} visualizations are presented alongside personalization suggestions with a rationale for manual implementation (\cref{fig:vignettes4} B). \textit{UIPulse} presents these suggestions through pop-up messages. For example, a message might state: ``\textit{Most used item. Suggestion: Place in a better position}''. 

    \textit{UIPulse} also presents cost-benefit messages with estimates regarding time savings and the duration of the personalization process (similar to mixed-initiative systems \citep{bunt2007understanding, bunt2009mixed}): ``\textit{We estimate that this personalization would require approximately four minutes, yielding a benefit of three seconds per visit}''. These estimates were not computed algorithmically but informed by our prior empirical observations of users performing comparable personalization tasks \citep{alves}.
    
    \item \textit{Visual Personalization Suggestions.} The vignettes display visual suggestions, pre-configured UI modifications that users can accept (\cref{fig:vignettes4} C), accompanied by a rationale and a cost-benefit message (e.g., ``\textit{We estimate that you will be 20\% faster with this shortcut. Click to confirm.}''). This approach draws on system-driven approaches \citep{kuhme1993user, jameson2007adaptive, Liu2024, wu2022reflow}, where UIs adapt automatically.

    \item \textit{Social Personalization Suggestions.} \textit{UIPulse} offers socially-informed suggestions (\cref{fig:vignettes4} D) in the form of community-based templates \citep{alves, 10.1145/3544549.3585668}, suggested according to behavioral patterns shared with similar users. This reflects a common content personalization practice, where social recommender systems leverage social information to guide tailored suggestions \citep{Hong27082025}. We introduced this set to explore participants' comfort with the data-sharing and analysis practices required to generate such recommendations. It also supported discussions on whether participants would find value in accessing aesthetic-related data and suggestions. An example is: ``\textit{30\% of our users are unhappy with the design. Click to import the following alternative}''.
\end{enumerate}

\subsubsection{Reflection Moments}
The vignettes support five reflection moments (\textit{Reflection 1--5} in \cref{fig:vignettesdiagram}), each typically consisting of three vignettes accompanied by questions about the data and the software. The first two vignettes of each moment ask participants to speculate about the personas' behaviors with that version of \textit{UIPulse}, using the personas as provocations to prompt critical reflection. The third vignette then shifts the focus to the participant, inviting them to clarify or expand on their own perspective.

Unlike the other three sets, set \textit{A} includes two reflection moments: the first focused on the general utility of interaction data, and the second on its potential for supporting UI personalization. \textit{Reflection 1} (following the \textit{Data-Only Visualization Dashboards} illustrations)  explores participants' views on the personas' context, concerns, and motivations for using \textit{UIPulse} (e.g., ``How do you think Chris can improve his/her interactions?''), followed by questions about participants' own attitudes toward monitoring digital interactions, data utility, and privacy (e.g., ``How would you feel about the security and privacy of your data?''). This moment is followed by a vignette illustrating a customization tool \citep{nebeling2013crowdadapt}. Immediately afterward, the second reflection moment shifts the discussion to personalization. It prompts participants to consider what data might be important or missing for supporting meaningful personalization, and how Alex and Chris could improve their interactions.

The remaining three moments (\textit{Reflection 3--5}) follow a similar structure. After each set, participants are invited to reflect on the personas' reactions to the \textit{UIPulse} features illustrated, how those features might influence data interpretation, and the perceived advantages and disadvantages compared to previous sets. In the last step of each moment, participants rate on four scales (1 to 5) the likelihood of (1) Alex and (2) Chris personalizing their UIs and (3) the benefit (4) and effort they expect of analyzing the data and personalizing with the features available in that set. We used these scales to encourage discussion rather than to analyze their numeric answers.

The last reflection moment (\textit{Reflection 5}) also explores potential privacy concerns related to social suggestions, along with general impressions of the study, particularly any perceived data gaps and alternatives for integrating interaction data into personalization.

\subsection{Participants and Recruitment}
We disseminated the study using university mailing lists, social media platforms, and by re-contacting individuals who had participated in previous studies on UI personalization. Given that the study relied on participants discussing fictional scenarios, perspectives from individuals with hands-on personalization experience were considered valuable for grounding the discussion. We aimed to include participants with and without personalization experience, although such experience was not used as an inclusion or exclusion criterion. To be eligible, participants had to be internet users for more than four hours per week. Participants received a €10 gift card in appreciation of their participation.

\begin{table}[h]
\centering
\begin{tabular}{c|c|c|c|c}
\toprule
\textbf{ID} & \textbf{Age} & \textbf{\begin{tabular}{@{} c @{}}UI Design\\ Expertise\end{tabular}} & \textbf{\begin{tabular}{@{} c @{}}Daily Digital\\ Activity (hours)\end{tabular}} & \textbf{\begin{tabular}{@{} c @{}}UI Pers.\\ Experience\end{tabular}} \\
\midrule
\textbf{P1} & 29 & \checkmark & 13 & — \\

\textbf{P2} & 29 & \checkmark & 13 & \checkmark\\

\textbf{P3} & 26 & \checkmark & 19 & —\\

\textbf{P4} & 27 & \checkmark & 12 & —\\

\textbf{P5} & 26 & \checkmark & 14 & \checkmark\\

\textbf{P6} & 48 & — & 11 & —\\

\textbf{P7} & 29 & — & 14 & —\\

\textbf{P8} & 60 & — & 10 & \checkmark\\

\textbf{P9} & 29 & — & 9 & —\\

\textbf{P10} & 32 & \checkmark & 13 & \checkmark\\

\textbf{P11} & 30 & — & 10 & —\\

\textbf{P12} & 36 & — & 10 & \checkmark\\
\bottomrule
\end{tabular}
\caption{Participant profiles by ID, age, expertise (professional UI design/development experience), daily digital activity (total hours a day using a smartphone and computer), and UI personalization experience, indicating prior participation in studies related to personalization.}
\label{table:participantss4}
\Description{The table lists 12 participants with their age, UI design or development expertise ("—" means no expertise), daily digital activity in hours, and prior UI personalization experience.}
\end{table}

We recruited participants with different expertise (education and technological proficiency) and demographic profiles (e.g., age and background). In total, 12 participants (P1 -- P12) concluded the study (\cref{table:participantss4}). Participants age varied between 26 and 60 years (33.4±10.3 years). Six identified themselves as men (50\%) and six as women (50\%). Participants usually spend 9.3±2.1 [7.5 -- 15] hours a day using a computer and 3±1.5 [1 -- 6] hours a day using a smartphone. They came from various educational and professional backgrounds, including healthcare, design, computer science, and economics. Six participants (50\%) were experts (defined as those with experience in UI design or development) while the other six (50\%) were regular users without such expertise. We judged data saturation to have been reached when two consecutive interviews yielded no new perspectives.

\subsection{Analysis}
Three researchers analyzed the audio-recorded interviews. We performed a thematic analysis \cite{braun2006using} following the first three stages outlined by \citet{HALCOMB200638}, which do not require complete transcription of interview recordings. First, one researcher took notes of relevant phrases, recurring ideas, and non-verbal cues during all interviews. After each interview, the researcher reviewed and expanded the notes. At the end of the study, the same researcher performed intelligent verbatim transcription by carefully listening to all audio recordings and selectively transcribing participant responses, excluding filler words or off-topic conversations. Given the semi-structured nature of the interviews, the resulting transcripts and notes were organized under thematic headings aligned with the interview questions, ensuring an accurate and coherent representation of participants' perspectives.

Next, in a group session, three researchers discussed and analyzed a sample of three (25\% of the dataset) interview transcripts. This initial review served to identify recurring themes, patterns, and insights. The team collaboratively constructed an affinity diagram to identify and connect emerging themes and group similar ideas. From this process, we developed a codebook that captured key themes and sub-themes reflecting participants' views on interaction data and personalization. One researcher used the codebook to code, reflect, and analyze all the transcripts. The research team iteratively discussed and refined themes for writing. We report our findings organized by these themes, supported by quotes and examples.

\section{Results}
In this section, we first describe participants' general response to engaging with interaction data (RQ1). We then outline the benefits and challenges of using interaction data to support personalization (RQ2), and conclude by presenting their expectations and requirements for integrating interaction data into personalization (RQ3).

%------------------------------------------------------------------------------------
%------------------------------------------------------------------------------------
%------------------------------------------------------------------------------------
%------------------------------------------------------------------------------------
\subsection{Engaging with Interaction Data}
As the study progressed, participants' understanding of how data can inform personalization decisions also evolved, leading to greater confidence in identifying UI changes. Their reactions to interaction data can be divided into three moments: (1) reflecting on the general utility, (2) personalization opportunities, and (3) personalization value.
%------------------------------------------------------------------------------------

\paragraph{General Attitude Towards Interaction Data} Before participants were debriefed on the study's focus on personalization, most were unable to connect the data with personalization. These challenges stemmed mainly from their initial agency perspective on UIs and personalization, focusing on using data to change their behavior and adapt themselves to the UIs rather than tailoring the UIs to their needs.

They were also skeptical about the possibility of monitoring their interactions. Five participants (P4, P5, P6, P7, P9) confessed a prior interest in monitoring their online behavior, primarily for better managing their digital routines. Others felt they already had a certain level of self-awareness regarding their interactions. Among the concerns, P10 believed that gaining deeper awareness could bring anxiety upon realizing their behavior was not optimized, and P8 expected disappointment when becoming aware of certain personal routines.

\paragraph{Identifying Personalization Opportunities} \textbf{With the personalization context, most participants (except P3, P8, and P9) intuitively and independently considered personalization opportunities without requiring system support for interpretation}, relying solely on the dashboards of the set \textit{A} and a brief description of a customization tool. The personalization opportunities they identified include ``\textit{creating shortcuts to increase task agility}'' (P4), ``\textit{opening websites automatically in the most used webpage}'' (P10), or ``\textit{better position the most clicked elements}'' (P11).

Click and scroll heatmaps emerged as the primary source for supporting the discovery of personalization opportunities, with participants highlighting their general preference for visual data dashboards: ``\textit{Click heatmaps are particularly interesting as they show where Alex clicks... also scroll heatmaps can support users in reorganizing the UI. If Alex constantly scrolls to the bottom, he may want to move the content to the top}'' (P7). They believed the presented data ``\textit{complements well}'' (P5) a personalization tool, highlighting the importance of ``\textit{complementing the visual perspective [e.g., heatmaps or charts] with a more quantitative one [e.g., tables]}'' (P3). The feedback of P12, with personalization experience, was encouraging: ``\textit{One of the difficulties I encountered in my previous customization experiences was: `What am I going to customize?' Heatmaps can be a good starting point for people to understand what UI elements they use most and should highlight}''.

Participants often contrasted their experiences with and without access to interaction data. For example, P5 emphasized how such data could have improved his past personalization efforts: ``\textit{I have previously used a personalization platform, and it was difficult to understand exactly the ideal arrangement for the UI elements. If I had access to these heatmaps, I could understand which buttons I access the most and distribute them more efficiently}''.

Nevertheless, other data can facilitate the process. P3 recommended including frequent navigation actions between different websites and information about multitasking (e.g., browser tabs frequently active simultaneously), which he expects to use to streamline workflows. P4 suggested incorporating zoom-in/out data to assess font-size adequacy, and P5 wanted to understand the time between clicks. Others, like P10, emphasized the need to interpret the semantics of interactions, noting that currently only the coordinates of the interactions (X and Y) are collected.

Some participants occasionally faced challenges identifying personalization opportunities, which they gradually overcame as they progressed through the vignettes. For example, they assumed that their routine actions or unconscious interactions with some UIs could make it challenging to identify issues or create moments for reflecting on potential improvements. P3 describes these interactions as part of a ``\textit{zombie mode}'' but assumes that accessing interaction data can make him reflect and analyze his actions while in this mode.

Other participants struggled to distinguish between the correct use of a UI (i.e., using UIs as intended by their designers) and the optimal use, which prioritizes efficiency. For instance, P2 focused more on finding interaction errors than UI improvements: ``\textit{Alex is using UIs correctly... so there is nothing to improve}''. Similarly, P12 found the data difficult to interpret and relied on textual suggestions to bridge the gap: ``\textit{Suggestions reduced the work and frustration of looking at data I do not understand... I should not have to understand heatmaps and suggestions guide me}''. Conversely, P2 found textual suggestions insufficient and requested visual comparisons: ``\textit{suggestions [B] are useless... Accessing a non-optimized and optimized way to use the UI would help to understand the benefits of suggestions}''.

\paragraph{Evaluating Personalization Opportunities} Before accessing the cost-benefit messages, participants struggled to identify the practical benefits of personalizing, as exemplified by P5: ``\textit{I can not estimate the benefit of this personalization. I would need data to be sure}''.

In that sense, \textbf{accessing cost-benefit metrics of personalization was valued by participants, with particular relevance for motivating the personalization step} (i.e., actually personalizing after identifying an opportunity). Participants noted that for people to personalize, they must first recognize it will lead to tangible improvements, as P10 highlighted: ``\textit{Suggestions would help me know what to change. Then, it would depend on how willing I would be to change. I personalized UIs before... but never felt it made a difference}''. Similarly, P5 emphasizes how estimates can facilitate personalization decisions by highlighting the value of each change: ``\textit{By knowing the benefit of personalizing each UI, I can select which UIs I personalize and their benefits would be higher}''.

Furthermore, in manual personalization contexts, cost-benefit metrics can help users ``\textit{plan}'' (P1) their personalization efforts by providing estimates of the time and effort required. According to participants, these metrics will also encourage users to accept visual suggestions: ``\textit{The probability of personalizing is a five [on a scale from one to five]. Alex knows how much time he will save and how much he will be faster}'' (P9).

Cost-benefit messages also encouraged participants to reflect on other details. For instance, participants double-checked the benefits by analyzing the time spent on each application (the screen time data of set \textit{A}), with more frequent usage correlating to a higher personalization probability. In that regard, participants suggested cost-benefit messages should account for usage frequency and present monthly time savings to facilitate benefits assessment.

Finally, these metrics also alleviated concerns about the effort and time required to analyze the data. Notably, while participants desired access to raw data, most preferred bypassing manual analysis in favor of these benefit estimates.

Their intentions to access the data were often related to mistrust in suggestions. Participants commented that a lack of trust in data and suggestions could hinder people's personalization intentions. Most expressed a desire to review the data themselves to validate and understand the suggestions, as P4 explained: ``\textit{I like to confirm the information and make sure I am making the right decision}''. They also highlighted that manual analysis can be vital for contextualizing behavior. For instance, P3 noted that users often engage with UI elements unconsciously (e.g., selecting text for highlighting it while reading), emphasizing that not all interactions are meaningful for personalization.

%------------------------------------------------------------------------------------
%------------------------------------------------------------------------------------
%------------------------------------------------------------------------------------
%------------------------------------------------------------------------------------
\subsection{Expected Benefits and Challenges of Interaction Data Collection and Analysis}
Participants anticipated benefits from accessing their own interaction data. Beyond informing UI personalization, they expected such data to enhance their awareness of digital behavior. They recognize the value of a reflection period to question their habits, which could reveal navigation in ``\textit{zombie mode}'' (P9), uncover unnoticed digital preferences or patterns, open navigation perspective to other opportunities, and ultimately support improved time management. Conversely, the most common challenge was the anticipated effort involved, with some participants expressing concerns about both the analysis and implementation efforts.

A central discussion point was privacy. Accessing interaction data can result in both benefits and challenges related to privacy. Five participants (P2, P3, P8, P11, P12) expressed privacy concerns. Most of these concerns, however, were not directly related to the specific solutions under discussion but to more general mistrusts. Participants described feeling that they are not the ``\textit{owners of their data}'', noting that ``\textit{data is already collected in background regardless of our decision}'' (P8), that they ``\textit{can not escape}'' (P8) data collection, and that they are ultimately ``\textit{limited to accept}'' (P5) it. This feeling of inevitability implies that most are still willing to engage in interaction data collection, with P2 assuming the need for storing data locally and P11 doing it only in applications not dealing with sensitive data. P3 was the only participant who assumed privacy as a barrier to installing \textit{UIPulse}, preferring personalization software that does not collect personal data: ``\textit{They could upload my data to their server without my consent}''.

Conversely, for many, accessing interaction data can positively affect privacy perceptions by bringing existing third-party data collection into the foreground, making it less opaque: ``\textit{At least I can see what companies are taking from me}'' (P3). P4 is another example: ``\textit{This data offers people more security, at least in knowing what is being collected and how it is being collected}''.

\textbf{For most, privacy is a secondary concern as long as the data proves beneficial (e.g., when suggestions clearly demonstrate added value) and people retain control over: (1) when and where data is collected, (2) what data is collected and stored, and (3) how it is used, including giving consent to its use}. In that sense, overall trust and perceptions of security increased as participants understood why data could be collected and had the opportunity to visualize it. Personalization suggestions were particularly reassuring, with P6 highlighting their importance in ``\textit{making clearer how the system uses the data}'' and showing that ``\textit{it is trying to help}''. P6 claims ``\textit{As long as there is trust and transparency, everything is fine}''. P7 summarized the general sentiment: ``\textit{If data is used to benefit me, security is a tertiary issue}''.

%Transparency can be crucial for highlighting benefits and increasing trust. 

On the negative side, \textbf{the social components generated greater concerns, mainly related to data disclosure fears}. Aesthetics-related suggestions incorporating social elements were well received; however, efficiency-related suggestions were met with more caution. P10 explained: ``\textit{When it is about aesthetics, it makes me feel more comfortable because it does not look like it is based on my data}''. Still, most participants are willing to share their data, albeit anonymously and untraceably (e.g., for average calculations).

%------------------------------------------------------------------------------------
%------------------------------------------------------------------------------------
%------------------------------------------------------------------------------------
%------------------------------------------------------------------------------------
\subsection{Integrating Interaction Data into Personalization Solutions}
Participants expect personalization to be beneficial, above all, to ensure more comfortable interactions with UIs, which involves having simpler UIs that support efficient navigation while keeping a pleasant aesthetic: ``\textit{When you use a UI frequently, you want to feel as comfortable as possible. It is worth changing to feel more comfortable, like creating a cozy home space}'' (P6). In this section, we report participants' responses to the four variations of \textit{UIPulse} that support personalization (RQ3).

\subsubsection{The Need for System Support}
While discussing set \textit{A} vignettes, although participants were able to identify personalization opportunities independently, most lacked the confidence to actively use the data for personalization. To increase the probability of personalizing, \textbf{participants expected system support regarding personalization decisions}: ``\textit{I would prefer a tool that presents this data together with personalization suggestions that I could select}'' (P5). The most clear case was P12, who highlighted the importance of system support in simplifying data analysis and implementation efforts: ``\textit{Personalization guidelines [i.e., instructions] would be great, even if I were ultimately the responsible for customizing... I would not need to analyze data. The instructions could include components like color or size that would allow me to reflect''}.

Following exposure to sets \textit{B} and \textit{C}, participants recognized the value of personalization suggestions, especially as the vignettes highlighted personalization opportunities they had missed, the benefits became clearer, and the effort required was reduced. Therefore, they reacted naturally and with excitement to the textual and visual suggestions: ``\textit{I like this (set B). It is really useful because it brings the two things (data and personalization) together. It makes sense because once the data is collected, it is easy to have these suggestions}'' (P7). For P9, suggestions can hold particular value for non-experts: ``\textit{We often know what we use or like in UIs, but it is difficult for a common user, like me, to imagine how UIs can be improved}''.

Participants frequently expressed their agreement with the suggestions as they read them (e.g., ``\textit{`Highlight the element.' Exactly... Next}'' (P6)), reflected on how implementing the suggestions can impact interactions, or considered alternative ways to implement them. They highlighted time-saving advantages by having their ``\textit{data immediately translated}'' (P11) by textual suggestions. However, they still expected support for also implementing suggestions, as highlighted by P10: ``\textit{It should have a suggestion and its preview}''.

This made visual suggestions the favorite concept for participants, as it was when visualizing them that they most fully appreciated the personalization concept. For P7, these suggestions represented a turning point in embracing personalization, noting ``\textit{This one (Set \textit{C}) brings it all together... it has to be this}''.

\subsubsection{Perceived Value and Limitations of Personalization Suggestions}
In addition to the benefits presented previously, participants' comments suggest that \textbf{accessing personalization suggestions can increase the likelihood of personalization, stimulate proactive personalization, and ultimately change their perspective on UIs and personalization.}

Suggestions, especially visual, can serve as triggers for personalizing: ``\textit{As Alex and Chris do not have to make the changes, just confirm, they will be more proactive and customize more}'' (P4). Participants outlined the advantages of visual suggestions compared with sets \textit{A} and \textit{B}, as P4 exemplified: ``\textit{These suggestions are a customization prompt. People will immediately feel prompted to use the system. While the other (set \textit{A}) satisfies curiosity about habits, these suggestions encourage reflecting on efficiency and personalization}''.

Similarly, participants' mindsets and intentions to personalize changed over the course of the study, particularly in response to the suggestions. For P4, ``\textit{suggestions trigger the first personalization step}'' and spark an intention of personalizing beyond the initial suggestion. Even visual suggestions, requiring only users' confirmation, can stimulate users to manually customize: ``\textit{Visual suggestions would be fantastic... They promote more active personalization, with users able to start personalizing themselves}'' (P12).

The most noticeable positive impact of suggestions was people's ``\textit{mindset change}'' (P6), resulting in the \textit{``acceptance that personalization is possible and improves UIs}'' (P6). Particularly, there was a change in people's vision towards their agency over UIs, as demonstrated by P7: ``\textit{Just as social media profiles reflect ourselves, why should not the websites we frequent also reflect a bit of who we are?}''.

Simultaneously, this mindset shift can encourage greater self-awareness (i.e., users' intentions to understand their own behaviors and interactions): ``\textit{I can start understanding the suggestions and increasing awareness about my use and, instead of waiting for suggestions, I can start making suggestions}'' (P12).

A mentality change was also visible when people speculated on extrapolating specific suggestions or opportunities to other contexts and UIs, as P8 mentioned: ``\textit{Alex will understand that he can change more things on every website he visits}''. Ultimately, participants began to understand the beneficial personalization opportunities without accessing the data or suggestions.

Conversely, they considered the downsides of visual suggestions to be minimal, as long as they maintained control over personalization decisions. These downsides included situations where participants did not relate to the suggestions, questioned their relevance, or found them visually unappealing. As P6 explained: ``\textit{There are opportunities I can identify independently and suggestions I do not like... but I have the option to accept or reject}''. Maintaining control implies having confirmation power, revert options, and the possibility of personalizing the visual suggestions further, as P9 expressed: ``\textit{On the popup there could be the option `Confirm' or `Keep Personalizing'}''.

\subsubsection{Control and Individualization as Prerequisites for Personalization}
Control is a key requirement for participants, including controlling the data collection process, data sharing, personalization decisions, and the personalized UI. This sense of control was also evident in participants' desire to access and review the data behind the suggestions presented earlier. Participants believed that being able to reflect on their data to make informed personalization decisions independently can foster a feeling of empowerment and make personalization feel more user-driven.

For most participants, the key includes balancing system-support with control, and feelings of ``\textit{individualization}'' (P11), ``\textit{belonging}'' (P7), and ``\textit{appropriation}'' (P12). Participants believe these emotions can emerge when users control the personalization process, and the resulting UI feels unique and shaped by their ``\textit{personal choice}'' (P7). For instance, they expect benefits from accessing UIs personalized by the community but highlighted they ``\textit{are not individual}'' (P7), although installing a community-based personalization template can also be a personal choice incorporating some degree of individualization and control: ``\textit{Multiple templates can be presented, and I decide. It would look more personal}'' (P7).

Participants' understanding of how data can inform personalization decisions evolved as the study progressed. This awareness evolution impacted their perceived sense of agency, and they gradually became more confident in identifying and proposing UI changes.

\subsubsection{Addressing the Disruption of User Habits}
Participants consistently reflected on how changing a UI would affect personal routines and the effort needed to adapt themselves to the personalized UIs. Some believed such changes could initially introduce inefficiencies, as P4 explained: ``\textit{You will not be faster initially because you are already used to it}''.

For participants, suggestions and cost-benefit metrics can be crucial in helping users feel confident in creating new routines. P5 believes people ``\textit{need to ensure personalizing will increase efficiency}''. P12 argues suggestions ``\textit{raise awareness of personalization benefits; and clarify the rationale for changing certain elements, which otherwise would conflict with users' mental map of UIs}''.

\subsubsection{The Role of Social-Informed Suggestions}
\textbf{Participants received social suggestions with caution, identifying opportunities mainly related to aesthetics}. For P5, having the ``\textit{community establishing the most pleasant aesthetic}'' makes sense. Participants also recognized the value of social suggestions for ``\textit{instantly [i.e., before data collection] accessing personalized UIs}'' (P11) and supporting users without ``\textit{patience}'' (P8) to personalize. Simultaneously, participants believe these suggestions can encourage users to experiment with different UIs, ultimately leading to adapting one's taste and greater awareness of UI elements susceptible to personalization.

In that sense, participants viewed social components as a form of ``\textit{social validation}'' (P8) for personalized UIs, helping to confirm if UIs are ``\textit{user-friendly}'' (P7). They also believed that social components could highlight community trends, though their opinions on their value vary. Some expressed skepticism, with P7 remarking, ``\textit{It seems designed for the sheeple}'', while others saw these influences more positively. For example, P6 likened them to fashion trends: ``\textit{It is like a nice piece of clothing that I would want to try}''.

Social components can still support the identification of personalization opportunities in several ways. They can draw attention to social tendencies, as illustrated by P4, who imagined a typical reaction: ``\textit{Everyone is using this, let me try it}''. They can also increase awareness of previously unnoticed UI elements, with P12, for instance, expressing surprise by noting, ``\textit{I did not know there was a box here}''. In addition, they can enable social comparisons, which can ultimately shape how people interpret their data, as P5 explained: ``\textit{If someone takes longer (to perform a task) compared to the community, it can indicate a problem}''.

\subsubsection{Personalization, Data, and Suggestions in Practice}
\textbf{To optimally support personalization, participants emphasized the need for a ``\textit{gradual personalization pipeline}'' (P1), which involves analyzing data, suggestions, and social trends}: ``\textit{I access the suggestions, then I need to know the general usefulness for people who already accepted them and for myself}'' (P1). P12 also connected these three steps, believing that accepting suggestions can create personalization habits, leading people to value more personalization and the social components.

A gradual personalization process can also work as a trust-building mechanism. For instance, P6 noted that people should go through progressive stages, testing each UI change and recognizing its benefits. By doing so, users can gain confidence in the system and its subsequent suggestions. Similarly, P10 emphasized that initial suggestions should be more subtle: ``\textit{The first suggestions need to have a high degree of certainty and be less visually disruptive}''. P10 suggests implementing suggestions as a ``\textit{tip of the day}'': ``\textit{I use a personalized UI for a few days, and when I get used to it, I can get a new personalization tip}''.

\section{Discussion}
This work studied opportunities to further empower users in personalization, not merely by offering options to change their UIs but by enabling informed decision-making. Through an interview study, we examined how putting interaction data at users' disposal can help them recognize and act on personalization opportunities, ultimately encouraging broader personalization adoption by enhancing its perceived value. In this section, we discuss our main findings, structured according to our research questions. Integrated within the discussion, we propose specific design considerations (DCs), which are then summarized in \cref{designconsiderations}.

%------------------------------------------------------------------------------------
%------------------------------------------------------------------------------------
%------------------------------------------------------------------------------------
%------------------------------------------------------------------------------------
\subsection{(RQ1) How do people engage with interaction data to identify and reflect on opportunities for UI personalization?}

Our findings suggest that enabling user access to interaction data enhances the value of personalization. People can identify personalization opportunities independently and by solely relying on data-only visualizations. Visual data representations, like click and scroll heatmaps, are particularly intuitive, enabling users to find opportunities for reordering UI elements and sections, hiding unused elements, or creating shortcuts for frequent operations. Conversely, quantitative data (e.g., usage metrics) is more useful for evaluating personalization benefits by enabling reflection on personal behavior and routines.

However, the main value of interaction data does not lie in supporting an initial unsupported and independent user analysis, but rather in enabling users to review and control their personalization decisions. Our findings suggest that simply exposing interaction habits for triggering personalization intentions, previously suggested by \citet{alves}, is insufficient. 
Future data-driven personalization systems should therefore focus on translating data into actions (e.g., visual personalization suggestions), with the results communicated in a convenient and user-friendly format to highlight the value of personalizing.

Interaction data should also be available for enabling users to review the data supporting suggestions. People often prefer to bypass data analysis and make personalization decisions based on cost-benefit metrics. However, they value reviewing interaction data for clarifying the motivations and benefits behind suggestions. This suggests user empowerment in personalization does not need to be necessarily linked to an unassisted setup, but rather to a system that enables a sense of control and informed decision-making.

Furthermore, systems should ensure persistent availability of interaction data to support proactive personalization. Our findings indicate that people can become more interested in actively personalizing as their confidence grows, a direct result of reflecting on suggestions and consulting interaction data.

These findings extend to users regardless of their expertise or UI personalization experience. However, the presentation of this data requires adaptation for non-experts, who often questioned specific visualizations (particularly scroll heatmaps) and struggled to interpret short-term efficiency gains. To address this, we recommend suggestion mechanisms to display long-term benefit metrics, as non-experts found cumulative savings easier to grasp and more motivating than one-session metrics.

Overall, access to interaction data showed promise to enhance existing personalization approaches, underscoring the importance of further exploring the concept through an in-the-wild study. Considering these insights, we formulate the following design considerations:
\begin{quote}
    (DC1) \textit{Empower User-Driven Personalization Decisions.}
\end{quote}
\begin{quote}
    (DC2) \textit{Use Complementary Data Representations.}
\end{quote}
\begin{quote}
    (DC3) \textit{Provide Meaningful Benefit Metrics.}
\end{quote}

%------------------------------------------------------------------------------------
%------------------------------------------------------------------------------------
%------------------------------------------------------------------------------------
%------------------------------------------------------------------------------------
\subsection{(RQ2) What are people's perceptions of the processes of collecting and analyzing interaction data for data-driven personalization, including its challenges and benefits?} \label{discussionbenefitsandcosts}

People recognize the value of personalization, and as long as the use of their data proves beneficial, most are willing to share it. We identified three interconnected mechanisms that future work must address to mitigate privacy concerns: controllability, transparency, and perceived utility. Accounting for these elements in the design of personalization systems may enhance users' trust in both user- and system-driven solutions. 

First, regarding controllability, design paradigms must shift to prioritize user agency \citep{mortier2014human, niu2025chat, rey2024understanding}. To mitigate the existing feelings of lack of control \citep{auxier2019americans}, future software should provide users with the ability to manage and curate their data (e.g., limiting collection to specific times and contexts).

Second, our findings suggest that visualizing collected data and linking it to personalization suggestions can enhance transparency and awareness of data usage. While prior research \citep{Sundar2010} suggests that awareness of data collection and privacy practices can increase distrust in system-driven personalization, we argue that this awareness can be fostered while preserving user trust. To support this, future solutions should focus not only on enabling data visualizations but also on clarifying how the data is being used, for instance, helping users understand how their data informs specific suggestions. This aligns with work on data-centric explanations, which shows that making decisions interpretable can increase system trustworthiness \citep{10.1145/3411764.3445736}.

Furthermore, our findings suggest that allowing users to review the data underlying personalization suggestions at their convenience can enhance their sense of control, potentially enhancing long-term engagement with the software. This complements prior research, which indicates that explanations are not always necessary, either because the effort to access them may outweigh their benefits \citep{10.1145/2166966.2166996}, or because users typically seek them only in response to specific triggers such as anomalous results or confusion \citep{10.2307/249487}.

Finally, future solutions should also focus on emphasizing the value and advantages of data collection, as there is a notable relationship between the benefits data provides and people's willingness to disclose it, often referred to as the personalization-privacy paradox \citep{XU201142, chellappa2005personalization}. We argue that this should be done by presenting personalization suggestions supported by relevant information, including justifications, data, and benefit estimates.

To address these privacy concerns, we recommend the following design consideration:
\begin{quote}
    (DC4) \textit{Communicate Data Usage Clearly.}
\end{quote}

%------------------------------------------------------------------------------------
%------------------------------------------------------------------------------------
%------------------------------------------------------------------------------------
%------------------------------------------------------------------------------------
\subsection{(RQ3) How do people envision personalization solutions enabling access to interaction data, and what support do they expect for identifying opportunities, making decisions, and implementing changes?}

While people can independently identify personalization opportunities, they still seek software support for identifying and implementing those opportunities. Next, we highlight key considerations for integrating visual suggestions into people's routines, offering insights for user- and system-driven personalization approaches.

\subsubsection{Designing Personalization Suggestions}
Personalization suggestions should not be an endpoint but work as a foundation or trigger for additional UI changes, allowing users to refine or expand upon the initial suggestions if desired. Personalization guidance can be crucial in alleviating the mental effort required to consider personalization details \citep{10.1145/3491102.3501931, alves}. Our results indicate that beyond guidance, people need a starting point for personalization, which visual suggestions can provide.

To this end, suggestions should have a clear rationale, and users should be able to explore their underlying data optionally. Previous efforts to provide rationale \citep{bunt2007supporting, bunt2009mixed, 10.1145/3411764.3445736} have focused on explaining algorithms or data usage, but participants either ignored or criticized these rationales for lacking personalized or graphical information. Our findings indicate that people value rationales that connect system suggestions with their personal data. They also suggest that if this connection is made, people can better reflect, make informed decisions, understand the benefits, grasp why specific data is collected, and become more aware of how to personalize. This increased personalization awareness can, in turn, strengthen their sense of agency over UIs, encourage a more critical perspective on new interfaces, help them identify similar opportunities in other contexts, and ultimately foster greater independence from system decisions in personalization.

To be meaningful, suggestions should also include cost-benefit metrics that account for design disruption. People value these metrics for planning the personalization and supporting their decision-making. However, they could be improved by also highlighting long-term benefits and capturing the disruption caused by changes, such as the time it takes users to become more efficient with the personalized UI compared to the original.

We also found that personalization suggestions can enhance users' proactivity (i.e., their willingness to personalize without being prompted by suggestions), indicating that suggestions can play an educational role. Future research should delve deeper into using suggestions to advise or guide users, fostering their awareness of key concepts and empowering them to take control of their UIs proactively.

Synthesizing these insights, we formulate the following design consideration:
\begin{quote}
    (DC5) \textit{Use Suggestions to Foster Proactivity.}
\end{quote}

%------------------------------------------------------------------------------------
\subsubsection{Introduction Strategies for Suggestions}
Our findings highlight the importance of introducing personalization gradually. A gradual approach can be vital for preventing abrupt routine changes (including lowering the learning curve of the personalized UI) and building user trust by starting with suggestions that are understandable, predictable, and clearly beneficial, such as adapting menus while maintaining spatial stability \citep{findlater09:design}.

To support effective gradualness, we propose that systems should allow users to engage with and decide on adaptations. This approach can strengthen the sense of control and foster trust by clearly demonstrating personalization benefits at each change. This approach is consistent with the principles of the slow computing philosophy, which emphasize reflection, pacing, and user agency in digital interactions \citep{kitchin2020slow}.

While prior work presents differing perspectives on whether users should be required to interact with adaptations or be allowed to ignore them \citep{findlater09:design}, we argue that user interaction is preferable, provided the trade-offs of obtrusiveness are carefully managed \citep{jameson2007adaptive}. Systems should ensure that suggestions are high-utility, contextually appropriate, and timed to avoid interrupting the user's primary tasks.

Furthermore, the personalization process does not need to be limited to one adaptation at a time. Future research should explore how many simultaneous UI changes can be made without overwhelming users or disrupting their mental models. Although individual adaptations can be helpful, implementing multiple individual adaptations in a short time frame risks confusing users and reducing their effectiveness. 

Based on these strategies, we derive the following design consideration:
\begin{quote}
    (DC6) \textit{Introduce Changes Gradually.}
\end{quote}

%------------------------------------------------------------------------------------
%------------------------------------------------------------------------------------
%------------------------------------------------------------------------------------
%------------------------------------------------------------------------------------

\begin{table*}[h!]
\centering
\begin{tabularx}{\textwidth}{p{0.25\textwidth} | X}
\toprule
\textbf{Design Consideration (DC)} & \textbf{Description} \\
\midrule

\textbf{(DC1) \mbox{Empower User-Driven} \mbox{Personalization Decisions}} &
Preserve user-driven personalization and control by providing system-initiated support through visual suggestions, while enabling optional customization and interaction data analysis.\\
\midrule

\textbf{(DC2) Use Complementary \mbox{Data Representations}} &
Offer both numerical (e.g., usage statistics, time-saving metrics) and visual representations (e.g., heatmaps) to support users in different decision-making processes, such as identifying personalization opportunities and evaluating their potential value.\\
\midrule

\textbf{(DC3) Provide Meaningful \mbox{Benefit} Metrics } &
Incorporate cost-benefit metrics that emphasize long-term benefits and frequency of use in benefit assessments.\\
\midrule

\textbf{(DC4) Communicate Data Usage Clearly} &
Provide interaction data visualizations and highlight its role in the personalization process (e.g., by linking data with suggestions).\\
\midrule

\textbf{(DC5) Use Suggestions to Foster Proactivity} &
Provide informative and educational personalization suggestions that can be a starting point or trigger for independent and proactive personalization.\\
\midrule

\textbf{(DC6) Introduce Changes \mbox{Gradually}} &
Introduce personalization changes progressively to prevent abrupt changes to users' routines or UI disruptions. Take advantage of this gradual approach to build trust by starting with suggestions that users are more likely to understand, deem relevant, and accept.\\

\bottomrule
\end{tabularx}

\caption{\label{table:implicationsSummary}Design considerations (DCs) for future user interface personalization with interaction data.}
\Description{Table with two columns (Design Consideration (DC) and Description) and six rows, each with a design consideration title and description.}
\end{table*}

\subsection{Design Considerations for Future User Interface Personalization}\label{designconsiderations}
This work represents an initial exploration to inform the design of systems that support users' personalization decisions through meaningful interaction data. \cref{table:implicationsSummary} summarizes our main design considerations to inform future work in this topic.

To translate these ideas into practice, we envision a tool with system-initiated design considerations where users can lead the personalization process through data visualization and user-driven personalization features. This system would offer visual suggestions, along with trade-offs highlighting long-term effort and time savings, optional data analysis, and further personalization options. Suggestions would be connected to interaction data, enabling users to make choices independently and proactively by gaining insight into the rationale behind the suggestions. UI elements would be personalized gradually, ensuring users' adaptation without disruption. The system would also give users control over the data collection context and communicate the long-term benefits, building trust and enhancing security perceptions. Such a system would advance existing personalization approaches, facilitating the identification of personalization opportunities and the ideation and implementation of solutions (two challenges of user-driven personalization \citep{alves, 10.1145/3491102.3501931}), while ensuring control, transparency, and understanding of system decisions beyond what existing system-driven or mixed-initiative personalization solutions offer \citep{bunt2007understanding, bunt2009mixed, Liu2024} and fostering users to engage with their data and personalize proactively.

%------------------------------------------------------------------------------------
%------------------------------------------------------------------------------------
%------------------------------------------------------------------------------------
%------------------------------------------------------------------------------------
\subsection{Limitations}
Using synthetic data and fictional personas to construct the design probes helped protect participants' privacy and ensured they felt comfortable participating in the study. However, this approach may limit deeper, more meaningful discussions about personalization opportunities, participants' digital routines, and real-world privacy concerns. While participants did situate their own use cases within the study context, future work could build on these findings to conduct an in-the-wild study to capture more authentic user experiences.

Additionally, the four scenarios and two personas reflect only a subset of possible system configurations and user identities. Other dimensions, such as system provenance or personas' abilities, cultural norms, and lived experience, could have generated additional insights into how people envision data practices in personalization. These dimensions were intentionally under-specified, as the personas served as conversational scaffolds to help participants project their own experiences, expectations, and concerns onto the scenarios, rather than reasoning about what a specific fictional user would do. Future work could extend these findings by including a broader and more diverse participant sample to support a wider range of projected experiences.

\section{Conclusion}
In this work, we investigated the benefits and challenges of a reflexive personalization approach where individuals are supported to reflect on their digital interaction data to identify personalization opportunities and personalize. In a semi-structured interview study, 12 participants engaged with design probes in the form of vignettes to reflect on various approaches to integrating interaction data into user-driven personalization, and to consider the extent of system support required to analyze, interpret, and act upon that data.

Our results show that, regardless of UI design expertise, people can autonomously identify personalization opportunities by using interaction data. Nevertheless, they favor system-initiated design decisions with visual suggestions that they can use as a starting point for their personalization. Access to interaction data can also enhance transparency, help users assess the benefits and drawbacks of personalization, and raise awareness of UI design and personalization features, which can be important for supporting an independent and proactive personalization process.

We expect this work will inspire future research on providing users with access to their own interaction data to support UI personalization. Such research should prioritize transparent and informative data-driven suggestions that are introduced gradually and designed to foster user proactivity, while ensuring users can access the underlying data to clarify and review these suggestions.

\begin{acks}
We thank the participants in our user study. This project was supported by Fundação para a Ciência e
a Tecnologia through LASIGE Research Unit, refs. UID/00408/2025, DOI \url{https://doi.org/10.54499/UID/00408/2025}, and SFRH/BD/146847/2019, and Project 41, HfPT: Health from Portugal, funded by the Portuguese Plano de Recuperação e Resiliência. \textit{Grammarly} and \textit{ChatGPT} were used to improve grammar, spelling, and clarity. The authors assume full responsibility for the content of this manuscript.
\end{acks}

\bibliographystyle{ACM-Reference-Format}
\bibliography{sample-base}

\appendix

\section{Sample Vignettes}\label{appendixbignettes}

\begin{figure*}[h]
    \centering
    \frame{\includegraphics[width=0.8\textwidth]{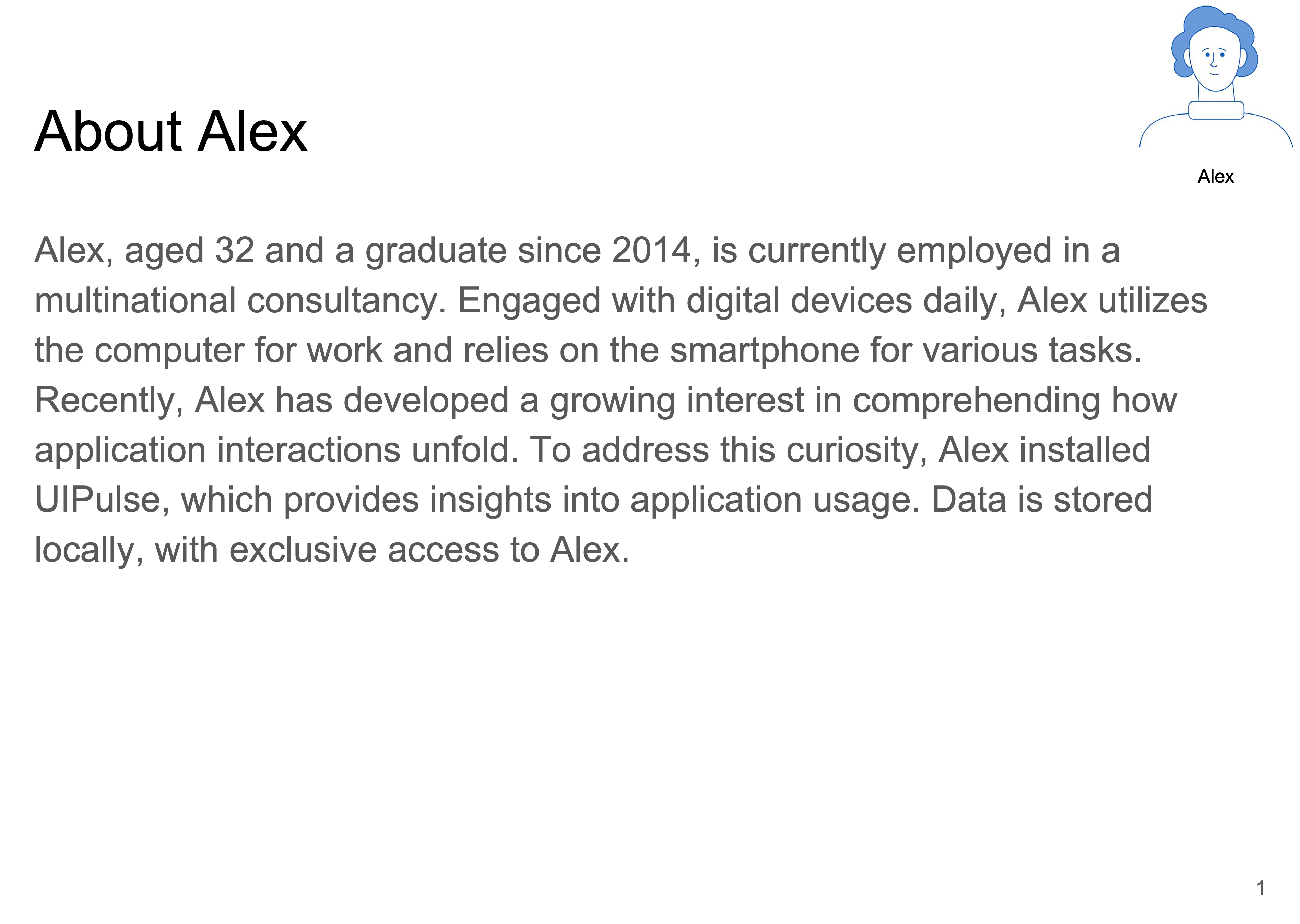}}
    \caption{Vignette 1: A description of the persona, Alex.}
    \Description{Vignette with a text describing Alex.}
    \label{fig:vignettesAlex}
\end{figure*}

\begin{figure*}[h]
    \centering
    \frame{\includegraphics[width=0.8\textwidth]{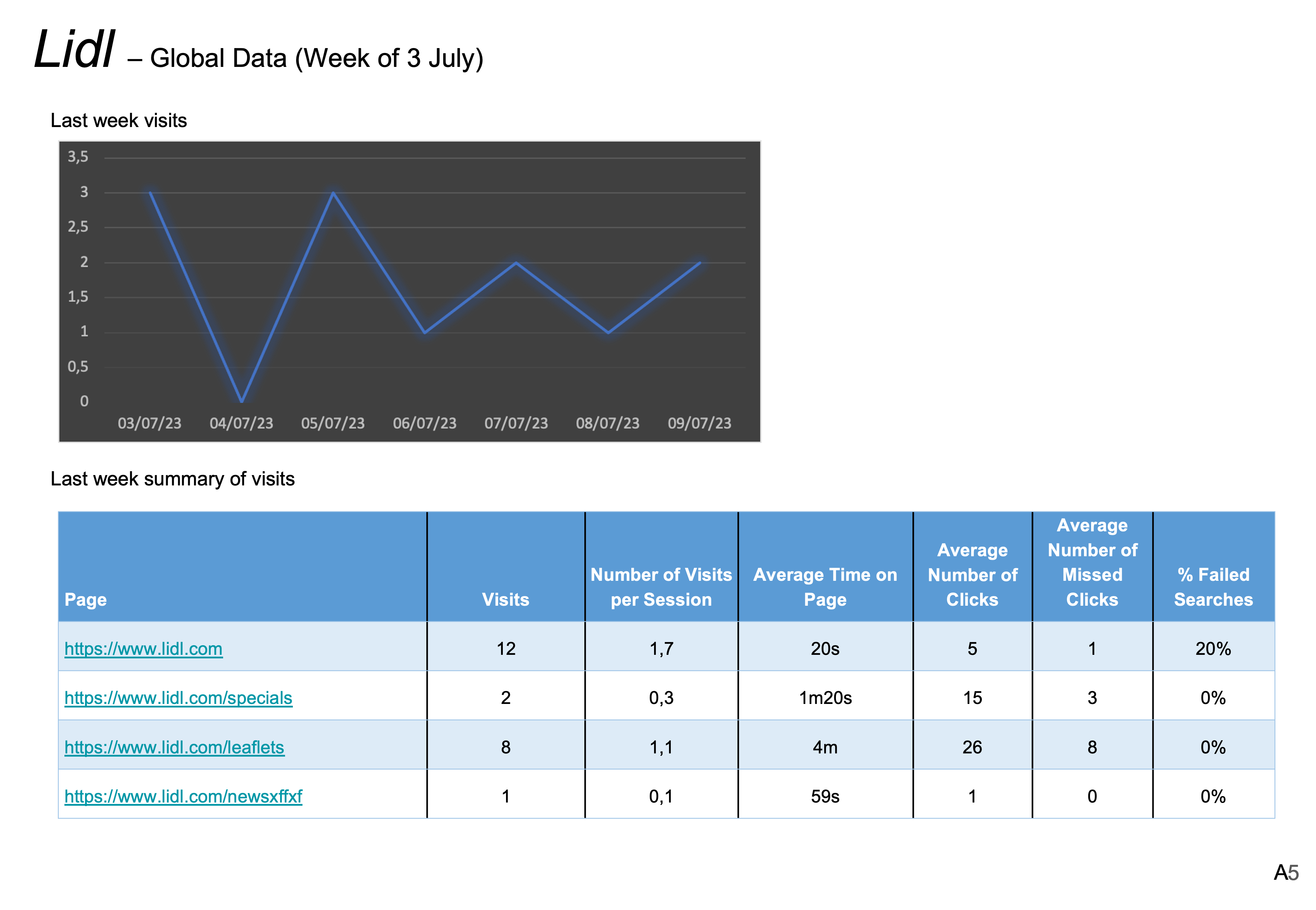}}
    \caption{Vignette A5: Illustration of last week's visits to a specific website.}
    \Description{Vignette with a chart and a table with information about last week's visits.}
    \label{fig:vignttesseta}
\end{figure*}

\begin{figure*}[h]
    \centering
    \frame{\includegraphics[width=0.8\textwidth]{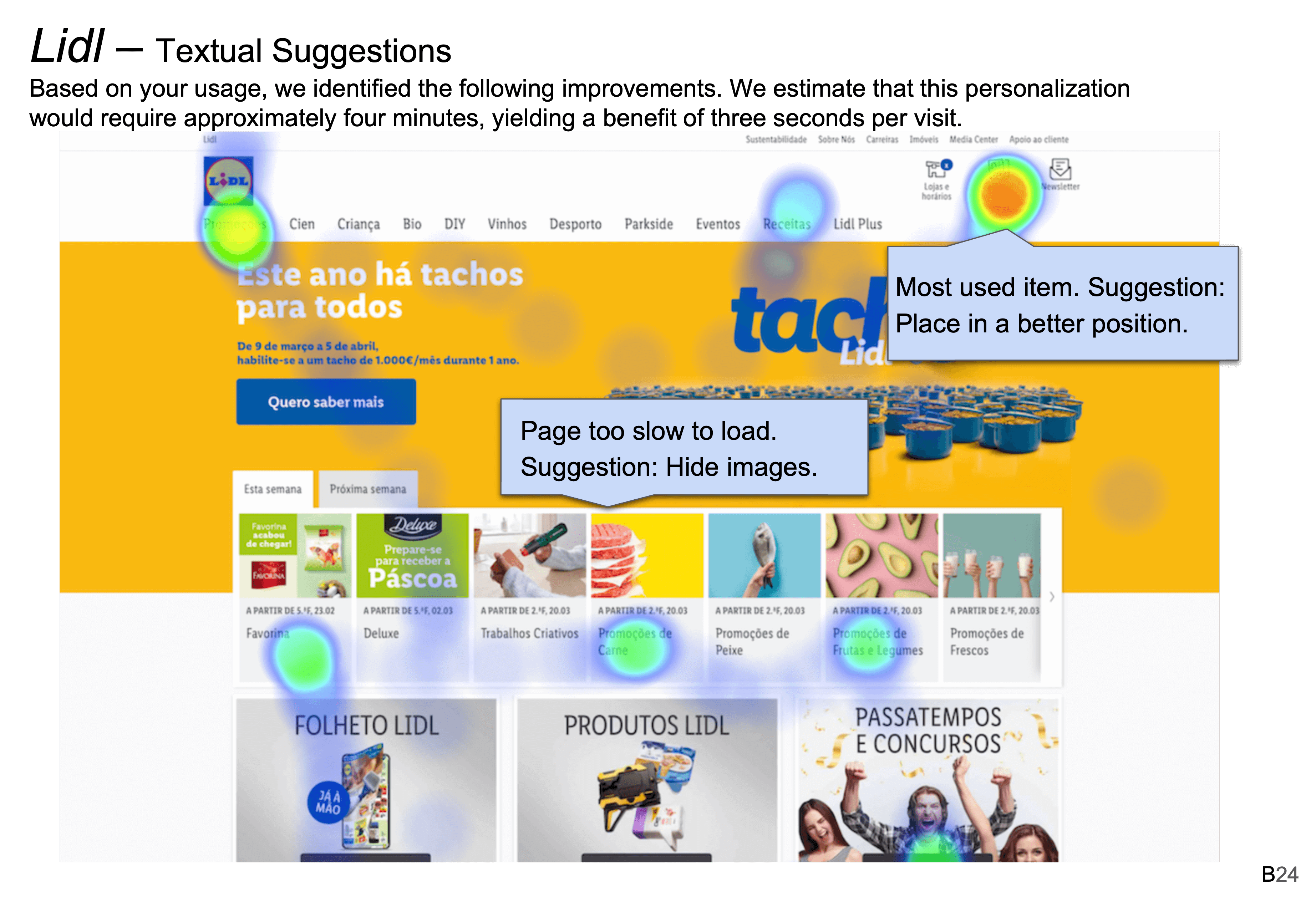}}
    \caption{Vignette B24: Illustration of textual suggestions.}
    \Description{Vignette with click heatmaps overlaid by textual suggestions.}
    \label{fig:vignttestextual}
\end{figure*}

\begin{figure*}[h]
    \centering
    \frame{\includegraphics[width=0.8\textwidth]{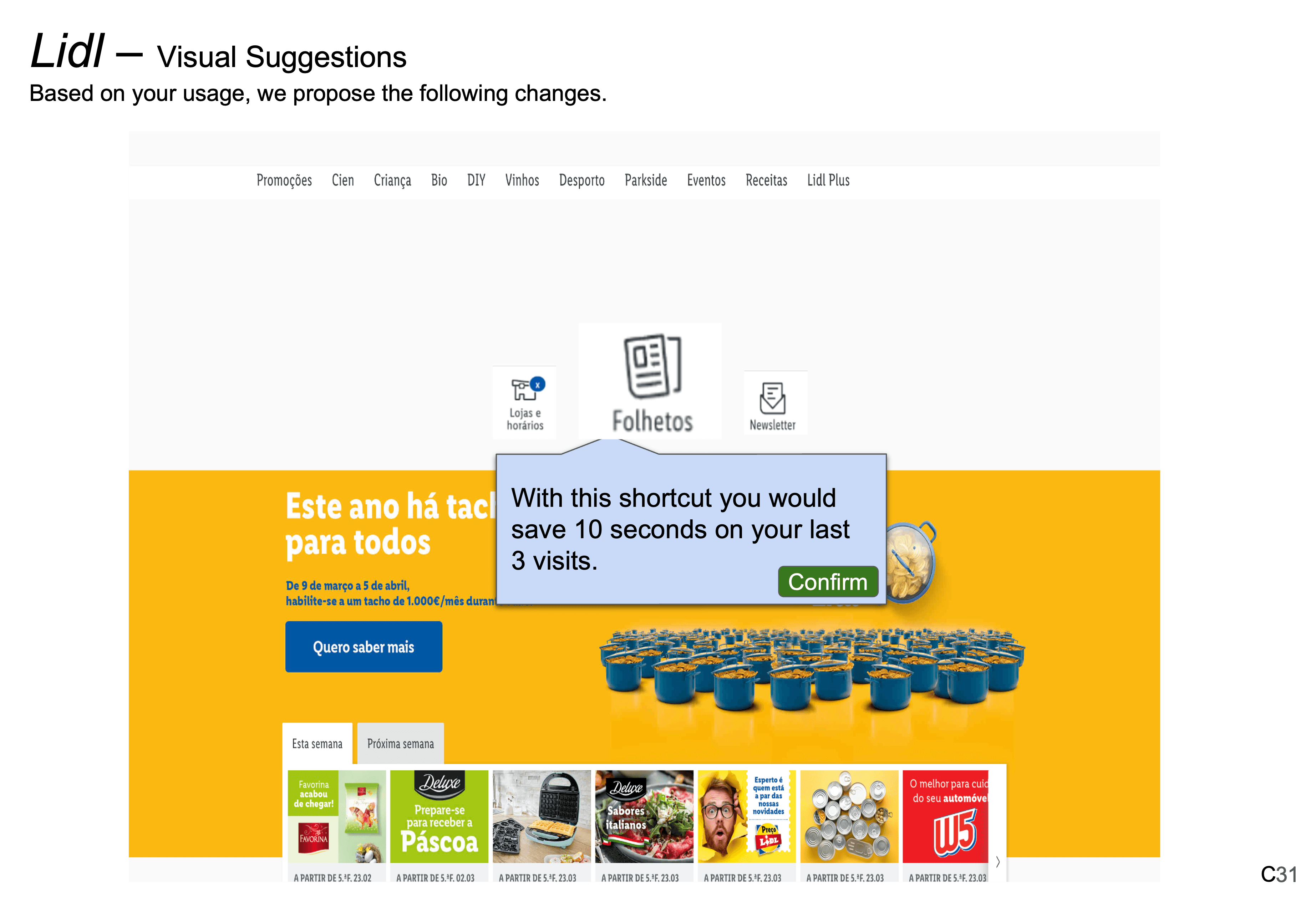}}
    \caption{Vignette C31: Illustration of visual suggestions.}
    \Description{Vignette with a personalized interface overlaid by textual information about the changes.}
    \label{fig:vigntettesvisual}
\end{figure*}

\begin{figure*}[h]
    \centering
    \frame{\includegraphics[width=0.8\textwidth]{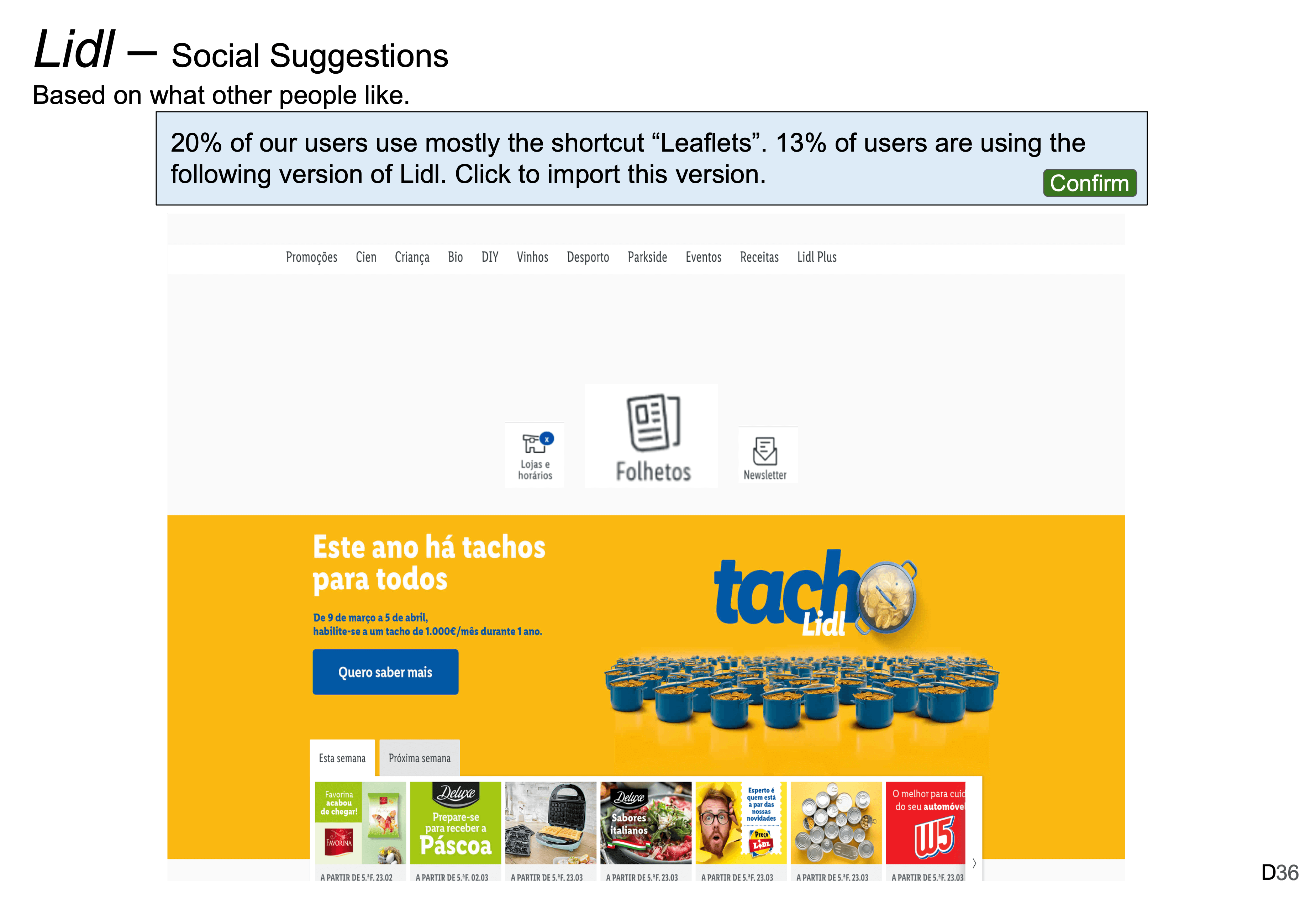}}
    \caption{Vignette D36: Illustration of social suggestions.}
    \Description{Vignette with a personalized interface and textual information about the changes based on social data.}
    \label{fig:vignettessocial}
\end{figure*}
    
\begin{figure*}[h]
    \centering
    \frame{\includegraphics[width=0.8\textwidth]{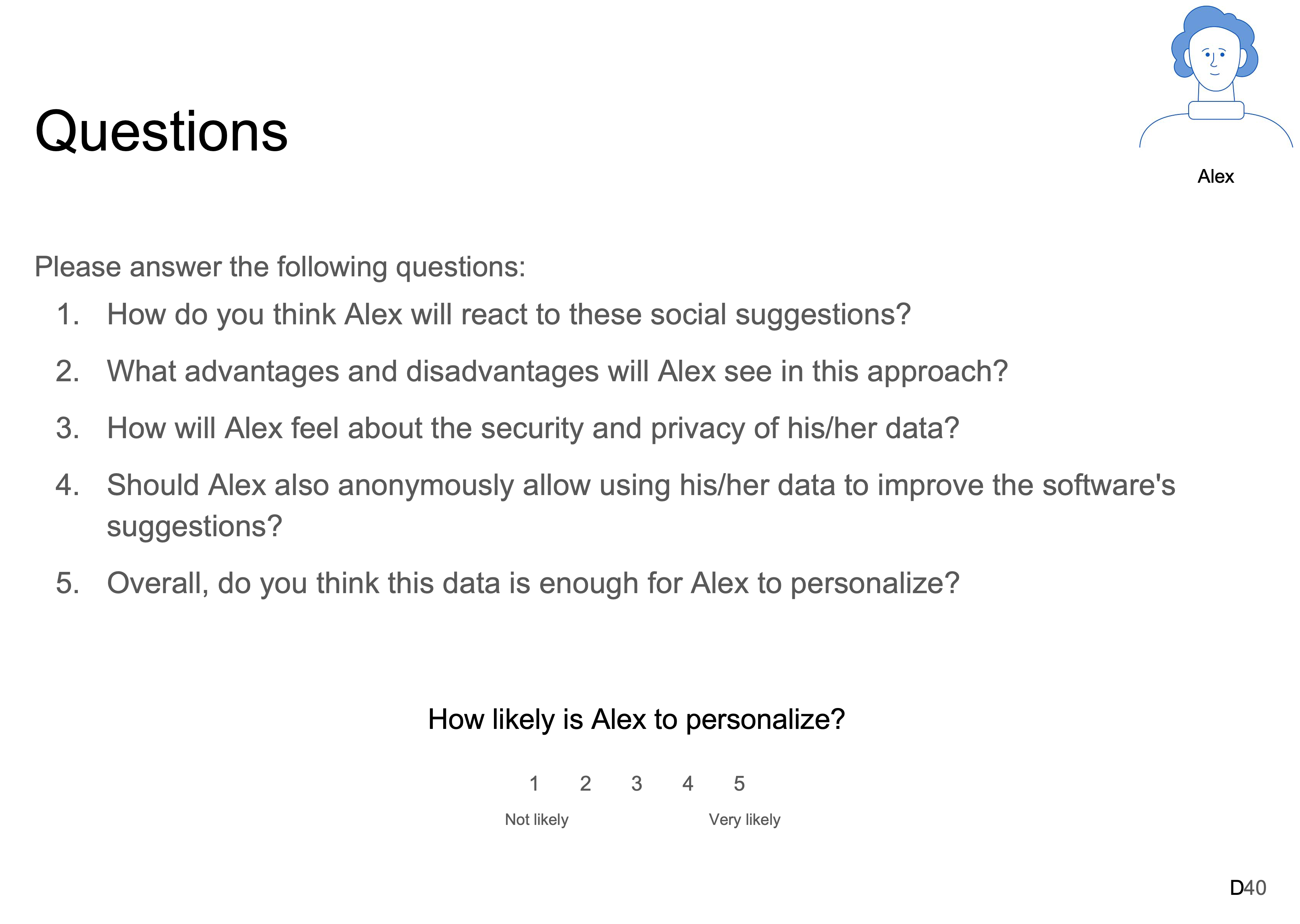}}
    \caption{Vignette D40: Questions used to guide the semi-structured interview discussion based on the scenario.}
    \Description{Vignette with a numbered list of questions.}
    \label{fig:vignettesquestions}
\end{figure*}

\end{document}